\documentclass[11pt]{article}
\title{Binary-state dynamics on complex networks: pair approximation and beyond}
\author{James P. Gleeson\\
\emph{MACSI, Department of Mathematics and Statistics,}\\
\emph{University of Limerick, Ireland.}\\
{james.gleeson@ul.ie}}
\def\kv{{\mathbf{{k}}}}
\usepackage{amsmath}
\usepackage{amsbsy}
\usepackage{amscd}
\usepackage{amsopn}
\usepackage{amstext}
\usepackage{amsxtra}
\usepackage{graphicx}
\usepackage{epsfig}

\begin{document}
\maketitle
\begin{abstract}
A wide class of binary-state dynamics on networks---including, for example, the voter model, the Bass diffusion model, and threshold models---can be described in terms of transition rates (spin-flip probabilities) that depend on the number of nearest neighbors in each of the two possible states. High-accuracy approximations for the emergent dynamics of such models on uncorrelated, infinite networks are given by recently-developed compartmental models or approximate master equations (AME). Pair approximations (PA) and mean-field theories can be systematically derived from the AME. We show that PA and AME solutions can coincide under certain circumstances, and numerical simulations confirm that PA is highly accurate in these cases. For monotone dynamics (where transitions out of one nodal state are impossible, e.g., SI disease-spread or Bass diffusion), PA and AME give identical results for the fraction of nodes in the infected (active) state for all time, provided the rate of infection depends linearly on the number of infected neighbors. In the more general non-monotone case, we derive a condition---that proves equivalent to a detailed balance condition on the dynamics---for PA and AME solutions to coincide in the limit $t \to \infty$. This permits bifurcation analysis, yielding explicit expressions for the critical (ferromagnetic/paramagnetic transition) point of such dynamics, closely analogous to the critical temperature of the Ising spin model. Finally, the AME for threshold models of propagation is shown to reduce to just two differential equations, and to give excellent agreement with numerical simulations. As part of this work, Octave/Matlab code for implementing and solving the differential equation systems is made available for download.
\end{abstract}
\section{Introduction} \label{sec:intro}
Binary-state dynamics on complex networks are frequently used as simple models of social interaction \cite{Castellano09,Castellano12a,Barratbook,Newmanbook}. Each node in a network is considered to be in one of two possible states (e.g., susceptible or infected, inactive or active) at each moment of time. Switching of nodes to the opposite state occurs stochastically, with probabilities that depend on the state of the updating node, and the states of its neighbors.
Examples include models for competition between two opinions in a network-based population, such as the voter model \cite{Liggettbook,Sood05} and the majority-vote model \cite{Liggettbook,deOliveira92,Pereira05}. Models for rumor propagation, adoption of new technologies, or the spreading of behavior are often based on generalizations of disease-spread \cite{PastorSatorras01,Hill10,Young09} or individual-threshold \cite{Granovetter78,Watts02,Centola07,Dodds12} dynamics. Recently, data from online social networks has been mined to help increase the understanding of how individuals are influenced by their network neighbors \cite{Romero11,VerSteeg11,Kleinbergbook,GonzalezBailon11,Bakshy11,Hodas12}; further insight is provided by experiments where the network topology and social interactions are carefully controlled \cite{Centola10}. Binary-state dynamics provide the simplest test for newly-developed models \cite{Shao09,PerezReche11}; interest is often focussed on the critical parameters where the behavior of the model changes qualitatively, as at the paramagnetic/ferromagnetic phase transition that occurs in the Ising spin model at the critical temperature \cite{Dorogovtsev02,Leone02,Galam82}.

All these models, and many others, can be considered as special cases of a general formulation for stochastic binary-state dynamics. Letting $k$ be the degree (number of network neighbors) of a node $X$ in an unweighted, undirected, uncorrelated, static network, we denote by $m$ the number of neighbors of $X$ who are in state 1, i.e., infected/active/spin-up, depending on the model context. If node $X$ itself is in state 0 (i.e., susceptible/inactive/spin-down) then the rate $F_{k,m}$ is defined by letting $F_{k,m} dt$  be the probability that $X$ will switch from state 0 to state 1 in an infinitesimally small time interval $dt$. On the other hand, if node $X$ is in state 1, then $R_{k,m} dt$ is the probability that node $X$ will switch to state 0 within the time interval $dt$. We call the functions $F_{k,m}$ and $R_{k,m}$ the \emph{infection rate} and \emph{recovery rate} respectively; collectively they are also called the \emph{transition rates} or \emph{spin-flip rates} of the dynamics. These rates depend (only) on the degree $k$ of node $X$, and the number $m$ of neighbors of $X$ who are in state 1; we also assume that the rates $F_{k,m}$ and $R_{k,m}$ do not vary in time. In Sec.~\ref{sec:derivation} (see Table~\ref{tabFR}) we show that many stochastic binary-state social interaction models  can be expressed in terms of suitable $F_{k,m}$ and $R_{k,m}$ functions \cite{Castellano12b,Vazquez08b}. Deterministic threshold models \cite{Granovetter78,Watts02,Centola07}---where nodes change state (become active) only when the number $m$ of active neighbors exceeds a pre-defined threshold level---may also be put into this form (see Sec.~\ref{sec:thresh}).

Approximations for macroscopic quantities of interest---such as the expected fraction of active nodes in the network at a given time---are needed in order to understand how the combination of  network topology (e.g., the degree distribution $P_k$) and the microscopic dynamics (defined by the rates $F_{k,m}$ and $R_{k,m}$) affect the emergent dynamics at macroscopic level. In the limit of infinite network size, systems of deterministic differential equations may be derived to describe such macroscopic quantities, at various levels of approximation. The most common analytical approach for dynamics on complex networks is \emph{mean-field (MF) theory}. Mean-field theories are typically derived under a number of assumptions, the most important of which for the current discussion is the assumed lack of \emph{dynamical correlations} \cite{footnote1}. Under this assumption, the neighbors of a node $A$ are considered to be active (i.e., in state 1) with a probability that is determined by the overall fraction of active nodes (of same degree) across the network as a whole. In particular, the probability of a neighbor of $A$ being active is assumed to be independent of the state of node $A$: thus there is assumed to be no correlation between the state of $A$ and the state of $A$'s neighbors.

Examples of  MF theories for binary-state dynamics are found in \cite{PastorSatorras01,Sood05, Castellano06,Barratbook}, and the validity of the assumptions of mean-field theories on networks are considered in detail in \cite{Gleeson12, Melnik11}. Mean-field theories, though relatively simple to derive and solve, are often quite inaccurate, especially on sparse networks (with low mean degree $z$) and near to critical points of the dynamics. Pair approximation (PA) theories improve upon the accuracy of MF theories---at the cost of extra complexity in the resulting system of differential equations---by including dynamical correlations at a pairwise level \cite{Eames02,Levin96,deOliveira93,Taylor12b,Vazquez08,Vazquez10,Schweitzer09,Dickman86}. In addition to the fraction of active nodes in each degree class (as in MF), PA theories account for the fraction of edges in the network that link active nodes to active nodes (called \emph{I-I edges} here), inactive nodes to inactive nodes (\emph{S-S edges}), and inactive to active nodes (\emph{S-I edges}). However, dynamical correlations beyond nearest neighbors are not captured by PA theories. If node $A$ is inactive, for example, the probability that each of its neighbors is in the active state is approximated in PA theory by the relative preponderance of S-I edges over S-S edges in the whole network. Moreover, each neighbor of $A$ is considered to be independent of every other neighbor of $A$. This means that if $m$ is the number  of $A$'s neighbors that are in the active state, $m$ is assumed to have a binomial distribution.

Higher-order accuracy, beyond PA level, has been obtained recently using compartmental models or \emph{approximate master equations (AME)} \cite{Marceau10,Lindquist11,Gleeson11,Petermann04}. These approaches consider the state of nodes and their immediate neighbors, generating large systems of differential equations, which better capture the (non-binomial) distributions of the number of active neighbors of updating nodes. The increased complexity of the differential equations systems gives improved accuracy over PA and MF, particularly near critical points of the dynamics \cite{Gleeson11}.

In this paper, we investigate the possibility of, for certain classes of dynamics, reducing the dimensionality of the full AME system to obtain a simpler set of differential equations, but without any loss of accuracy.
Ideally, the reduced-dimension system would permit analytical solutions, but even if closed-form solutions are not possible, the smaller system is less computationally expensive to solve.
We restrict our attention to static networks, but note that AME approaches have also been successfully used for models where the network structure co-evolves with the dynamics \cite{Marceau10,Durrett12,Taylor12a}. We also assume here that the networks are generated by the configuration model \cite{Bender78,Bollobas80,Newmanbook}, so having zero clustering (transitivity) in the infinite-size limit. These simplifications enable us to focus on the types of dynamics for which the AME can be exactly reduced to a simpler (e.g., pair approximation) system; we anticipate that complicating factors such as non-zero clustering can be studied in subsequent work, for example by employing the models of \cite{Newman09,Gleeson09,Hackett11,HebertDufresne10}.

The remainder of the paper is organized as follows. In Sec.~\ref{sec:derivation} we give examples of binary-state dynamics and in Sec.~\ref{sec:approx} we briefly review the approximate master equation approach of \cite{Gleeson11}, showing how the equations of PA and MF theories can be derived as systematic approximations of the full AME. In Sec.~\ref{sec:mono} the AME and PA solutions are shown to give  identical results for certain macroscopic quantities within the class of dynamics we call \emph{monotone}. In Sec.~\ref{sec:general} a more general class of dynamics (corresponding to equilibrium spin systems obeying detailed balance) is shown to have AME and PA solutions that are equal in the steady-state limit  $t\to \infty$, but not at finite $t$. Focussing on equilibrium models with up-down symmetry, this result enables us to derive---in Sec.~\ref{sec:spin}---an analytical result for the critical point of such dynamics, i.e., the bifurcation point marking the phase transition from paramagnetic (disordered) to ferromagnetic (ordered) behavior, as in the Ising model on complex networks \cite{Leone02,Dorogovtsev02}. Finally, in Sec.~\ref{sec:thresh}, we show that (monotone) threshold models may be accurately approximated using an extended version of the AME, and that the large system of equations may be reduced to just two differential equations to determine the expected fraction of active nodes. Details of some mathematical derivations are contained in the Appendices.

\section{Examples of binary-state dynamics} \label{sec:derivation}
\begin{figure}
\centering
\epsfig{figure=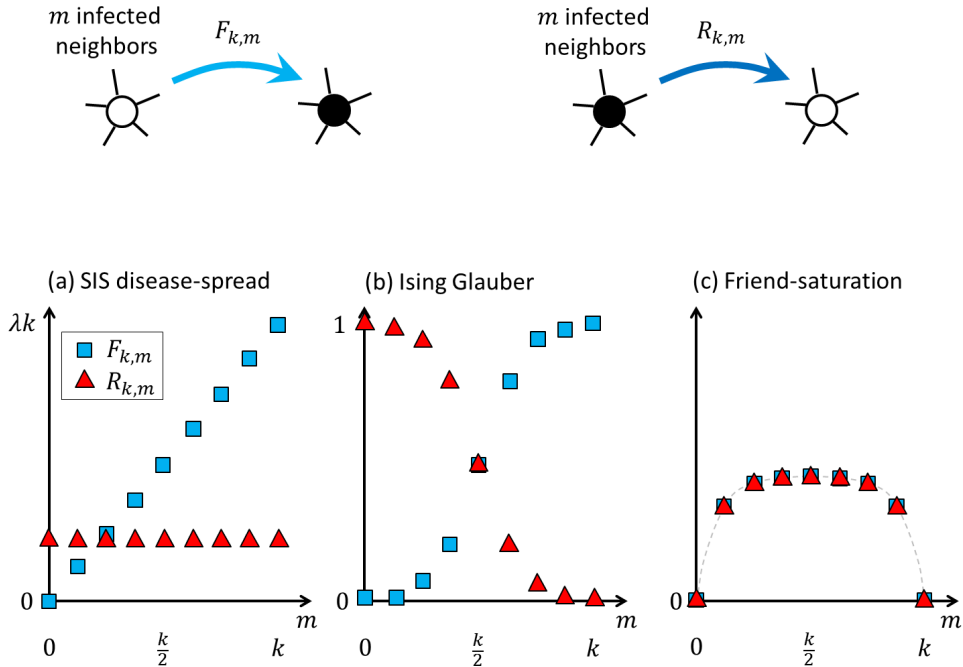,width=13cm}
\caption{Schematic of the infection rate $F_{k,m}$ and recovery rate $R_{k,m}$  as defined in the text. Here---and in Figs.~\ref{fig_schematic} and \ref{fig2_inkscape}---white nodes are susceptible/inactive/spin-down and black nodes are infected/active/spin-up. Examples of the dependence of the rates on the number $m$ of infected neighbors: (a) susceptible-infected-susceptible disease spread model, (b) Glauber dynamics for the Ising model, (c) a stylized ``friend-saturation'' model. Case (c) is motivated by the data analysis  of social contagion for Digg stories in Fig.~4 of \cite{VerSteeg11}, where the probability of, for example, becoming infected (voting on a Digg story) initially increases with the number of infected neighbors, but then saturates, and eventually decreases, giving a function that is roughly symmetric about $m=k/2$.} \label{fig:FRschematic}
\end{figure}

\begin{table}
\begin{center}

\begin{tabular}{|c|c|c|c|c|}
\hline
Process & Infection rate & Recovery rate  &    Symmetric             & Equilibrium\\
  or model      &    $F_{k,m}$       &     $R_{k,m}$        &model & model\\
\hline\hline
SIS  & $\lambda m$ & $\mu$   &  No  & No\\ \hline 
SI   & $\lambda m$ &   0     &  No  & No\\ \hline
Bass & $c+d m $    &   0    & No    & No\\ \hline
Kirman & $c_1+d m $  & $c_2+ d(k-m) $ & No & No \\ \hline
voter  & $\frac{m}{k}$ &  $\frac{k-m}{k}$ & Yes & No\\ \hline  
$\begin{array}{c}\text{link}\\\text{update}\\\text{voter}\end{array} $ & $\frac{m}{z}$ & $\frac{k-m}{z} $ & Yes & No \\ \hline
$\begin{array}{c}\text{nonlinear}\\\text{voter on}\\\text{4-regular}\\\text{random}\\\text{graph \cite{Schweitzer09}} \end{array} $& $\left\{\begin{array}{l}
                                F_{4,0}=0\\ F_{4,1}=\alpha_1\\F_{4,2}=\alpha_2\\F_{4,3}=1-\alpha_2\\ F_{4,4}=1-\alpha_1 \end{array}\right.$ &
                                $\left\{\begin{array}{l}
                                R_{4,0}=1-\alpha_1 \\ R_{4,1}=1-\alpha_2\\R_{4,2}=\alpha_2\\R_{4,3}=\alpha_1\\ R_{4,4}=0 \end{array}\right.$ & Yes & No \\ \hline
$\begin{array}{c}\text{language}\\\text{model}\end{array} $ & $s\left(\frac{m}{k}\right)^\alpha$ & $(1-s)\left(\frac{k-m}{k}\right)^\alpha$ & Yes, if $s=\frac{1}{2}$ & No \\ \hline
  $\begin{array}{c}\text{majority-}\\\text{vote}\end{array} $& $\left\{ \begin{array}{cc}
Q & \text{ if }m<k/2 \\
1/2 & \text{ if }m=k/2\\
1-Q & \text{ if }m>k/2 \\
\end{array} \right. $&  $\left\{ \begin{array}{cc}
1-Q & \text{ if }m<k/2 \\
1/2 & \text{ if }m=k/2\\
Q & \text{ if }m>k/2 \\
\end{array} \right. $ & Yes & No\\ \hline
 $\begin{array}{c}\text{Ising}\\\text{Glauber}\end{array} $ & $\frac{1}{ 1+ \exp\left(\frac{2 J}{T}(k-2m)\right)}$ & $\frac{\exp\left(\frac{2 J}{T}(k-2m)\right)}{ 1+ \exp\left(\frac{2 J}{T}(k-2m)\right)}$ & Yes & Yes\\ \hline
$\begin{array}{c}\text{Ising}\\\text{Metropolis}\end{array} $& $\left\{ \begin{array}{cc}
e^{\frac{2 J}{T}(2m-k)} & \text{ if }m<k/2 \\
1 & \text{ if }m\ge k/2
\end{array} \right. $  &  $\left\{ \begin{array}{cc}
1 & \text{ if }m\le k/2 \\
e^{\frac{2 J}{T}(k-2m)} & \text{ if }m>k/2
\end{array} \right. $ & Yes & Yes\\ \hline
threshold & $\left\{ \begin{array}{cc}
0 & \text{ if }m< M_\mathbf{k} \\
1 & \text{ if }m\ge M_\mathbf{k}
\end{array} \right. $ & 0 & No & No\\ \hline
 \hline
\end{tabular}
\end{center}
\caption{Infection and recovery rates for some examples of binary-state dynamics on networks: $k$ is the node's degree, $m$ is its number of infected neighbors, $z$ is the mean degree $\left<k\right>=\sum_k k P_k$ of the network. The parameters of the various models are described in Sec.~\ref{sec:derivation}. Symmetric models are those with rates that obey condition (\ref{T2}); equilibrium models obey condition (\ref{T}). }
\label{tabFR}
\end{table}
Here we briefly examine some of the binary-state models that can be described using transition rates $F_{k,m}$ and $R_{k,m}$; see Table~\ref{tabFR} and the schematic Fig.~\ref{fig:FRschematic}.
Note that in many of the models described here
 a single randomly-chosen node is updated at each time step, for which the time increment is $dt=1/N$, where $N$ is the number of nodes in the network (and we take the $N\to \infty$ limit). In such a model, $F_{k,m}$ is the probability that a spin-down node, when selected for updating, flips to spin-up, while $R_{k,m}$ is the corresponding probability of a node flipping from spin-up to spin-down \cite{footnote2}.

The susceptible-infected-susceptible (SIS) disease spread model \cite{Baileybook,Andersonbook,Harris74} assumes that infected individuals transmit disease to each of their network neighbors at a rate $\lambda$. Thus if a susceptible node has $m$ infected neighbors, the probability of it remaining susceptible for a time interval $dt$ is $(1-\lambda \, dt)^m$. Therefore its probability of infection is $1-(1-\lambda \, dt)^m$, and in the limit $dt \to 0$ this is $\lambda m dt$, giving the $F_{k,m}$ rate for SIS dynamics in Table~\ref{tabFR}. The recovery rate in the SIS model is a constant $\mu$, and in the case $\mu=0$ recovery is impossible; this special case is called the SI model.

The Bass model \cite{Bass69,Dodds04,Watts07} for diffusion of technological innovations is an extension of the SI model, and may be similarly specialized to a network context. In the Bass model, nodes move from the susceptible (non-adopted) state to the infected (adopted) state with rate $F_{k,m}=c+d m$. The parameter $c$ represents individual action, independent of the influence of neighbors, while the parameters $d$  quantifies the ``herding behavior'', whereby individuals copy the actions of their neighbors: in this case, by adopting the innovation when $m$ of their neighbors have already done so. Transitions from the adopted state to the non-adopted state are not permitted, so $R_{k,m}\equiv 0$.
Kirman's ant colony model \cite{Kirman93,Alfarano05,Gontis12} for choice between stock market trading strategies has similar herding effects, but with switching permitted between both states (strategies). A node with $m$ of its $k$ neighbors infected has $k-m$ neighbors in the susceptible state, so the rate for transitions of such a node from infected to susceptible is $R_{k,m}=c_2 + d (k-m)$. Note that mean-field theory (see Eq.~(\ref{MF}) below) for the complete-graph case $P_k=\delta_{k,N}$---where all nodes are connected to all other nodes---gives the standard population-level version of the Bass and Kirman models \cite{Bass69,Kirman93}.

In the voter model \cite{Liggettbook,Sood05}, a node is chosen at random in each time step ($dt=1/N$) and it adopts the same state of one of its (randomly-chosen) neighbors. If $m$ of the node's $k$ neighbors are in the active state, the node thus becomes active with probability $m/k$, and becomes inactive with probability $(k-m)/k$, these being the fractions of its neighbors in the respective states. Many variants of the voter model have been studied \cite{Castellano09,Barratbook}. In the link update voter model, for example, an edge (rather than a node) is chosen at random in each time step, and one of the end-nodes of the chosen edge then copies the state of the other. Nonlinear voter models and other variants have been examined by a number of authors \cite{Molofsky99,Abrams03,Vazquez08b,Schweitzer09}. In \cite{Schweitzer09} for example, the current state of the node is incorporated into the probabilities for switching state, lending an effective inertia to the dynamics: see the transition rates for this case (on $4$-regular random graphs) in Table~\ref{tabFR}. The language model of \cite{Abrams03} has been examined in detail in \cite{Vazquez10}; in this model the two states represent the primary language choice of a person (node), and the  probability of switching states is proportional to the fraction of speakers in the locality, raised to the power $\alpha$, multiplied by  the status parameter $s$ or $1-s$ of the respective language ($s=1/2$ for the symmetric case of equal-status languages, $\alpha=1$ then gives the voter model).

Many models of opinion dynamics are based on the classical Ising model of magnetic spins. Here each node is in either the spin-up or spin-down state and transitions occur according to dynamical rules which minimise the Hamiltonian of ferromagnetic interactions \cite{Barratbook,Krapivskybook}. Letting $T$ represent the temperature of the heat bath, and $J$ the ferromagnetic interaction parameter, the transition rates for Glauber \cite{Glauber63} and Metropolis \cite{Metropolis53} dynamics are shown in Table~\ref{tabFR}.
The majority-vote model \cite{Liggettbook,deOliveira92} is an non-equilibrium spin model, with spins tending to align with the local neighborhood majority, but with a noise parameter $Q$ giving the probability of misalignment.

As a final example, we consider threshold models \cite{Granovetter78,Watts02,Centola07,Gleeson07}, which are used to model diffusion of fads, collective action, or adoption of innovations \cite{Young09}. Each node has a (frozen) threshold level, which may depend on the degree
of the node. In each time step, a fraction $dt$ of the $N$ nodes in the network are updated. When updated, a node compares the number $m$ of its active neighbors to its threshold and activates (with probability 1) if $m$ is greater than or equal to its threshold.
Similar to the Bass model of innovation diffusion, usually no recovery is permitted in this model  \cite{Young09, Kleinbergbook} (but see \cite{Dodds12} for a recent extension including recovery). Note that a coordination game, modelling the diffusion of new behavior through a network, can also be written as a threshold model of this type \cite{Morris00,Kleinbergbook,Montanari10}. Threshold models, and the extension of the AME required to describe them, are considered in Sec.~\ref{sec:thresh} below.

\section{Approximate master equations, pair approximations, and mean-field theory} \label{sec:approx}

For completeness, we here briefly recapitulate the approximate master equations (AME) introduced in \cite{Gleeson11}; these were derived by generalizing the approach used in \cite{Marceau10,Lindquist11} for SIS dynamics, see Appendix~\ref{appA} for details.
Let  $s_{k,m}(t)$ (resp. $i_{k,m}(t)$) be the fraction of $k$-degree nodes that are susceptible (resp. infected) at time $t$, and have $m$ infected neighbors. Then the fraction $\rho_k(t)$ of $k$-degree nodes that are infected at time $t$ is given by
$
\rho_k(t) = \sum_{m=0}^k i_{k,m} = 1 - \sum_{m=0}^k s_{k,m}, 
$
and the fraction of infected nodes in the whole network is found by summing over all $k$-classes:
$
\rho(t) = \left<\rho_k(t)\right>\equiv \sum_k P_k \, \rho_k(t).
$
(Recall $P_k$ is the degree distribution, i.e., the probability that a randomly-chosen node has $k$ neighbors).

The approximate master equations for the evolution of $s_{k,m}(t)$ and  $i_{k,m}(t)$ are \cite{Gleeson11}:
\begin{align}
\frac{d}{d t}s_{k,m} &= -F_{k,m} s_{k,m} + R_{k,m} i_{k,m} - \beta^s (k-m) s_{k,m} + \beta^s(k-m+1) s_{k,m-1} \nonumber\\
 &\hspace{1cm}- \gamma^s m s_{k,m} + \gamma^s(m+1) s_{k,m+1}, \label{seqns}\\
\frac{d}{d t}i_{k,m} &= -R_{k,m} i_{k,m} + F_{k,m} s_{k,m} - \beta^i (k-m) i_{k,m} +\beta^i(k-m+1) i_{k,m-1}\nonumber\\
 & \hspace{1cm}- \gamma^i m i_{k,m} + \gamma^i(m+1) i_{k,m+1}, \label{ieqns}
\end{align}
for each $m$ in the range $0,\ldots,k$, and for each $k$-class in the network, with $\beta^s = \frac{\left<\sum_{m=0}^k (k-m) F_{k,m} \,s_{k,m}\right>}{\left< \sum_{m=0}^k (k-m) s_{k,m}\right>} $,
$\gamma^s =  \frac{\left<\sum_{m=0}^k (k-m) R_{k,m} \,i_{k,m}\right>}{\left< \sum_{m=0}^k (k-m) i_{k,m}\right>}  $, $\beta^i = \frac{\left< \sum_{m=0}^k m\, F_{k,m} \,s_{k,m}\right>}{\left< \sum_{m=0}^k m\, s_{k,m}\right>}$, and $\gamma^i =  \frac{\left< \sum_{m=0}^k m\, R_{k,m} \,i_{k,m}\right>}{\left< \sum_{m=0}^k m\, i_{k,m}\right>}$.
Equations (\ref{seqns}) and (\ref{ieqns}), with the time-dependent rates $\beta^s$, $\gamma^s$, $\beta^i$ and $\gamma^i$ (defined as nonlinear functions of $s_{k,m}$ and $i_{k,m}$), form a closed system of deterministic equations which can be solved numerically using standard methods; Octave/Matlab scripts are available from the author's webpage \cite{AME_solve}. Assuming a randomly-chosen fraction $\rho(0)$ of nodes are initially infected,  the initial conditions are
$
s_{k,m}(0)=\left( 1- \rho(0) \right) B_{k,m}(\rho(0))$, $i_{k,m}(0) = \rho(0) B_{k,m}(\rho(0)), 
$
where $B_{k,m}(q)$ denotes the binomial factor
\begin{equation}
B_{k,m}(q)=\binom{k}{m} q^m (1-q)^{k-m}. \label{binomialB}
\end{equation}

 For dynamics on a general network, with non-empty degree classes from $k=0$ up to a cutoff $k_\text{max}$, the number of differential equations in the system (\ref{seqns})--(\ref{ieqns}) is $(k_\text{max}+1)(k_{\text{max}}+2)$, and so grows with the square of the largest degree.
Some approximation is therefore necessary if it is desirable to reduce the AME to a lower-dimensional  system. One possibility is to consider the parameters $p_k(t)$ (resp. $q_k(t)$), defined as the probability that  a randomly-chosen neighbor of a susceptible (resp. infected) $k$-degree node is infected at time $t$. Noting that $p_k(t)$ can be expressed in terms of $s_{k,m}$ as $\sum_{m=0}^k m s_{k,m}/\sum_{m=0}^k k s_{k,m}$,
an evolution equation for $p_k$ may be derived by multiplying equation (\ref{seqns}) by $m$ and summing over $m$. The right-hand-side of the resulting equation contains higher moments of $s_{k,m}$, so a closure approximation is needed to proceed. If we make the ansatz that $s_{k,m}$ and $i_{k,m}$ are proportional to binomial distributions:
$
s_{k,m}\approx (1-\rho_k)\,B_{k,m}(p_k)$, $i_{k,m} \approx \rho_k\, B_{k,m}(q_k), 
$, we obtain the \emph{pair approximation (PA)}, consisting of the $3k_\text{max}+1$ differential equations:
\begin{eqnarray}
\frac{d}{d t}\rho_k &=& -\rho_k \sum_{m=0}^k R_{k,m} B_{k,m}(q_k) + (1-\rho_k)\sum_{m=0}^k F_{k,m}B_{k,m}(p_k),\nonumber\\
\frac{d}{d t}p_k &=& \sum_{m=0}^k \left[ p_k - \frac{m}{k}\right]\left[ F_{k,m} B_{k,m}(p_k) - \frac{\rho_k}{1-\rho_k} R_{k,m} B_{k,m}(q_k)\right] + {\beta^s}(1-p_k)-{\gamma^s}p_k,\nonumber\\
\frac{d}{d t}q_k &=& \sum_{m=0}^k \left[ q_k - \frac{m}{k}\right]\left[ R_{k,m} B_{k,m}(q_k) - \frac{1-\rho_k}{\rho_k} F_{k,m} B_{k,m}(p_k)\right] + {\beta^i}(1-q_k)-{\gamma^i}q_k ,\label{PA}
\end{eqnarray}
for each $k$-class. The rates here are given by  inserting the binomial ansatz into the general formulas, so that ${\beta^s}$, for example, is 
$\left<(1-\rho_k)\sum_m (k-m) F_{k,m}B_{k,m}(p_k)\right>/\left<(1-\rho_k)k(1-p_k)\right>$;
 initial conditions are $\rho_k(0)=p_k(0)=q_k(0)=\rho(0)$.

 A cruder, \emph{mean-field (MF)}, approximation results from replacing both $p_k$ and $q_k$ with $\omega$:
$
s_{k,m}\approx (1-\rho_k)\,B_{k,m}(\omega)$, $i_{k,m} \approx \rho_k\, B_{k,m}(\omega), 
$
where $\omega=\left< \frac{k}{z} \rho_k\right>$ is the probability that one end of a randomly-chosen edge is infected, and $z=\left<k\right>$ is the mean degree of the network.
Using this ansatz in the master equations yields a closed system of $k_\text{max}+1$ differential equations for the fraction $\rho_k$ of infected $k$-degree nodes:
\begin{equation}
\frac{d}{d t}\rho_k= -\rho_k \sum_{m=0}^k R_{k,m} B_{k,m}(\omega) + (1-\rho_k)\sum_{m=0}^k F_{k,m}B_{k,m}(\omega), \label{MF}
\end{equation}
with $\rho_k(0)=\rho(0)$.

In \cite{Gleeson11} we showed that the AME system (\ref{1}) generally gives improved accuracy over the approximations (\ref{PA}) and (\ref{MF}). Moreover, the general equations (\ref{PA}) and (\ref{MF}) for pair approximation and mean-field theory reduce to previously-studied equations when specific rates $F_{k,m}$ and $R_{k,m}$ are selected from Table~\ref{tabFR}. In this paper we concentrate on finding dynamics for which the AME system can be reduced exactly to lower-dimensional systems, for instance to the PA equations.

\section{Monotone dynamics} \label{sec:mono}

We first consider the case where $R_{k,m}=0$ for all $k$ and $m$, with $F_{k,m}$ nonzero for at least some arguments. The zero recovery rate means it is impossible for nodes to switch from  the infected state to the susceptible state and so we call this type of dynamics \emph{monotone} \cite{footnote3}. In this case the $s_{k,m}$ equations of the AME system are decoupled from the $i_{k,m}$ equations, with Eq.~(\ref{seqns}) reducing to
\begin{equation}
\frac{d}{dt} s_{k,m} = - F_{k,m}s_{k,m} - \beta^s (k-m)s_{k,m} + \beta^s(k-m+1)s_{k,m-1}, \label{eq:4}
\end{equation}
and the fraction of infected nodes is given by $\rho(t)=1-\sum_k P_k\sum_{m=0}^k  s_{k,m}(t) $. Thus several important macroscopic quantities (though not all) can be found by solving Eq.~(\ref{eq:4}) for $s_{k,m}(t)$, without considering the $i_{k,m}(t)$ variables.

Now consider whether or not the pair approximation ansatz $s_{k,m}(t) = (1-\rho_k(t))B_{k,m}(p_k(t))$---that was used to derive Eqs.~(\ref{PA})---could be an   exact solution of Eq.~(\ref{eq:4}). Inserting the PA ansatz into Eq.~(\ref{eq:4}) and dividing by $(1-\rho_k)B_{k,m}(p_k)$ gives the condition
\begin{equation}
-\frac{1}{1-\rho_k} \frac{d \rho_k}{d t} + \left( \frac{m}{p_k}-\frac{k-m}{1-p_k}\right) \frac{d p_k}{d t } = -F_{k,m} -\beta^s(k-m)+ \beta^s \frac{1-p_k}{p_k} m \label{eq:5*},
\end{equation}
where the identities
\begin{equation}
\frac{d}{d p} B_{k,m}(p) = \left(\frac{m}{p}-\frac{k-m}{1-p}\right) B_{k,m}(p)
\end{equation}
and
\begin{equation}
(k-m+1)B_{k,m}(p) = \frac{1-p}{p} m B_{k,m}(p)
\end{equation}
have been respectively utilized on the left-hand side and right-hand side of Eq.~(\ref{eq:5*}).

Equation~(\ref{eq:5*}) can be viewed as a condition on the forms of the infection rate $F_{k,m}$ for which the PA ansatz on $s_{k,m}$ is an exact solution of the corresponding approximate master equation. Note that all terms in Eq.~(\ref{eq:5*}) are linear in $m$, so $F_{k,m}$ is necessarily of the form
\begin{equation}
F_{k,m} = c_k + d_k m \label{lin},
\end{equation}
for (possibly $k$-dependent) constants $c_k$ and $d_k$.
Using this form in (\ref{PA}) confirms that the solutions of the AME for $s_{k,m}$ are given by the PA ansatz, with $\rho_k$ and $p_k$ being the solutions of the reduced system
\begin{align}
\frac{d}{dt}\rho_k &= (1-\rho_k)(c_k+k p_k d_k) \nonumber\\
\frac{d}{dt}p_k &= - d_k p_k (1-p_k) + \overline{\beta}^s(1-p_k),\label{eq:5}
\end{align}
where $\overline{\beta}^s = \left(\sum_k P_k(1-\rho_k) k (1-p_k)(c_k+(k-1)p_k d_k)\right)/ \left(\sum_k P_k(1-\rho_k) k (1-p_k)\right)$.
For the special case of $z$-regular random graphs or Bethe lattices (i.e., $P_k=\delta_{k,z}$ for integer $z$), system (\ref{eq:5}) can be solved analytically to give the explicit solution for the infected fraction as
\begin{equation}
\rho(t) = 1- (1-\rho(0)) e^{-(c+z d)t}\left[ 1- \frac{(1-\rho(0))d(z-2)}{c+d(z-2)}\left( 1-e^{-(c+d(z-2))t}\right)\right]^{-\frac{z}{z-2}}. \label{eq:6}
\end{equation}

\begin{figure}
\centering
\epsfig{figure=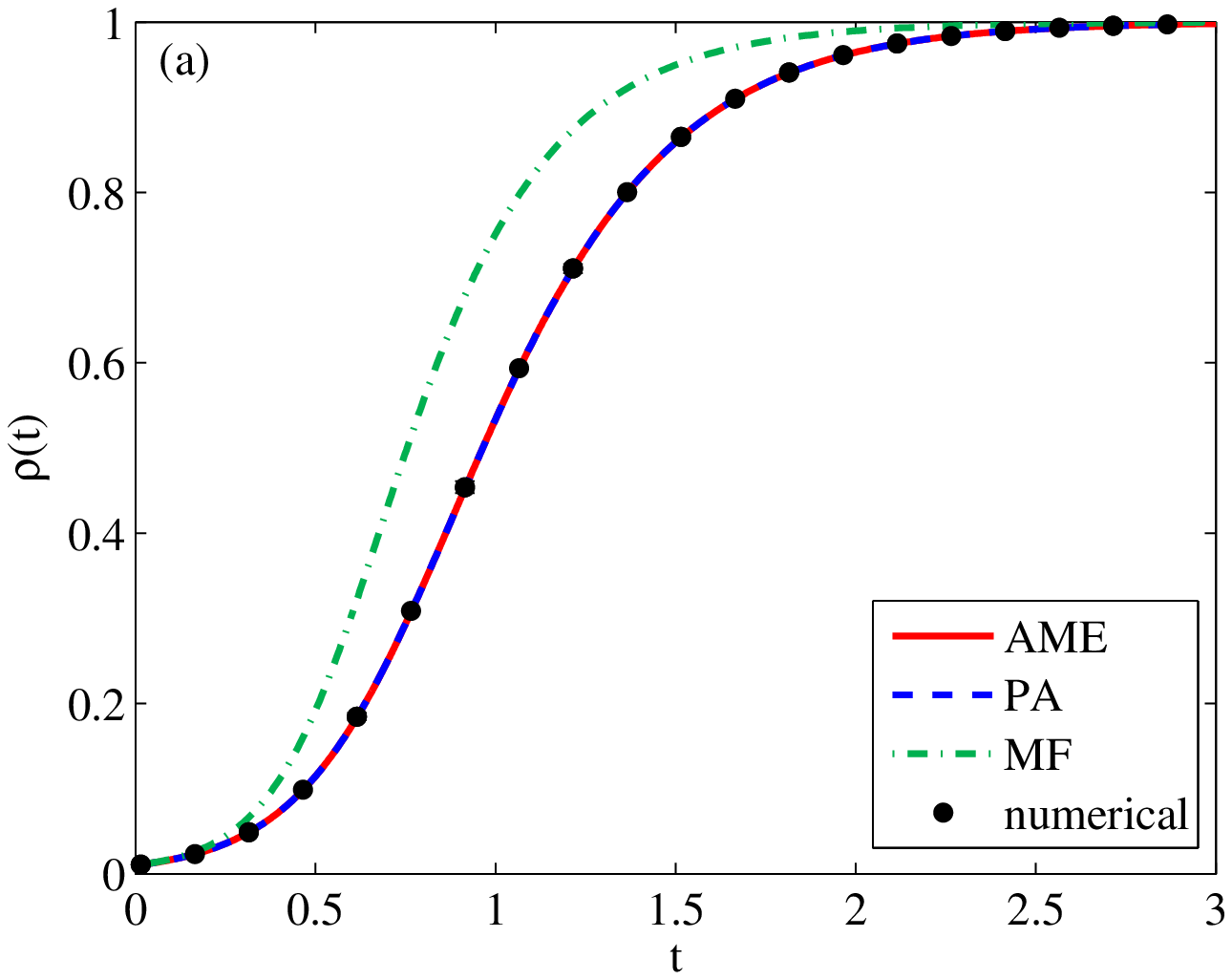,width=8cm}
\epsfig{figure=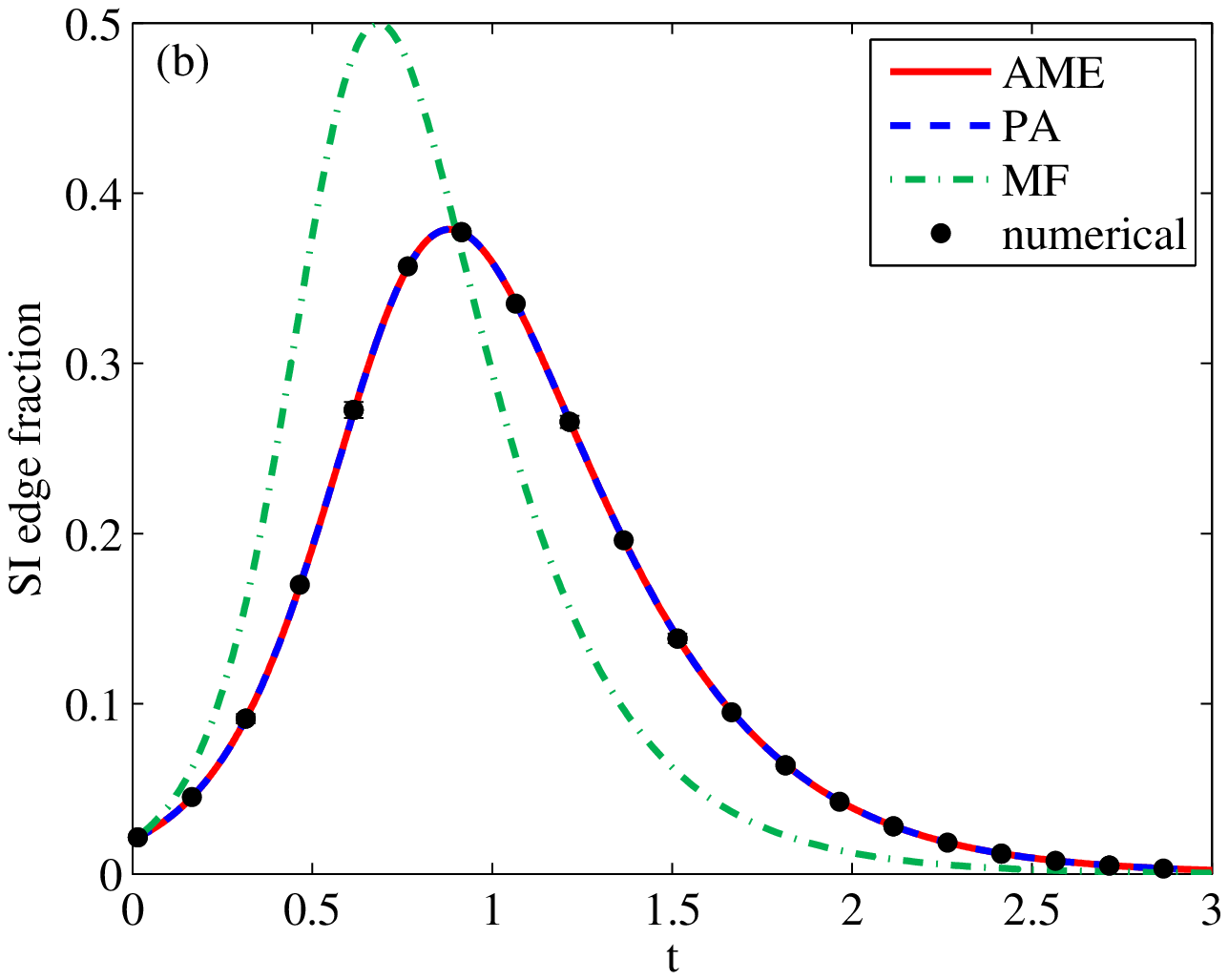,width=8cm}
\caption{ Susceptible-infected (SI) dynamics (with transmission rate $\lambda=1$) on a truncated scale-free network, with degree distribution given by $P_k \propto k^{-2.5}$ for degrees in the range $3\le k \le 20$, with $P_k=0$ otherwise. The initial infected fraction is $\rho(0)=10^{-2}$. Here, and in all subsequent Figures, symbols show the results of numerical simulations on networks of size $N=10^5$, using time step $dt=10^{-4}$. Results are averages over 24 realizations; error bars indicate mean $\pm$ one standard deviation (but note error bars are smaller than the symbols in this example).}\label{fig:SI}
\end{figure}
In Figure~\ref{fig:SI}  we show results for SI dynamics (see Table~\ref{tabFR}), which are monotone, and of form (\ref{lin}) with $c_k= 0$ and $d_k = \lambda$ for all $k$. Note the exact match between the AME and PA solutions for $\rho(t)$ and for the fraction of $S$-$I$ edges (see Eqs.~(\ref{star}) and (\ref{eq:55})) in the network, and the excellent agreement with numerical simulation on networks of size $N=10^5$ \cite{footnote4}.

\begin{figure}
\centering
\epsfig{figure=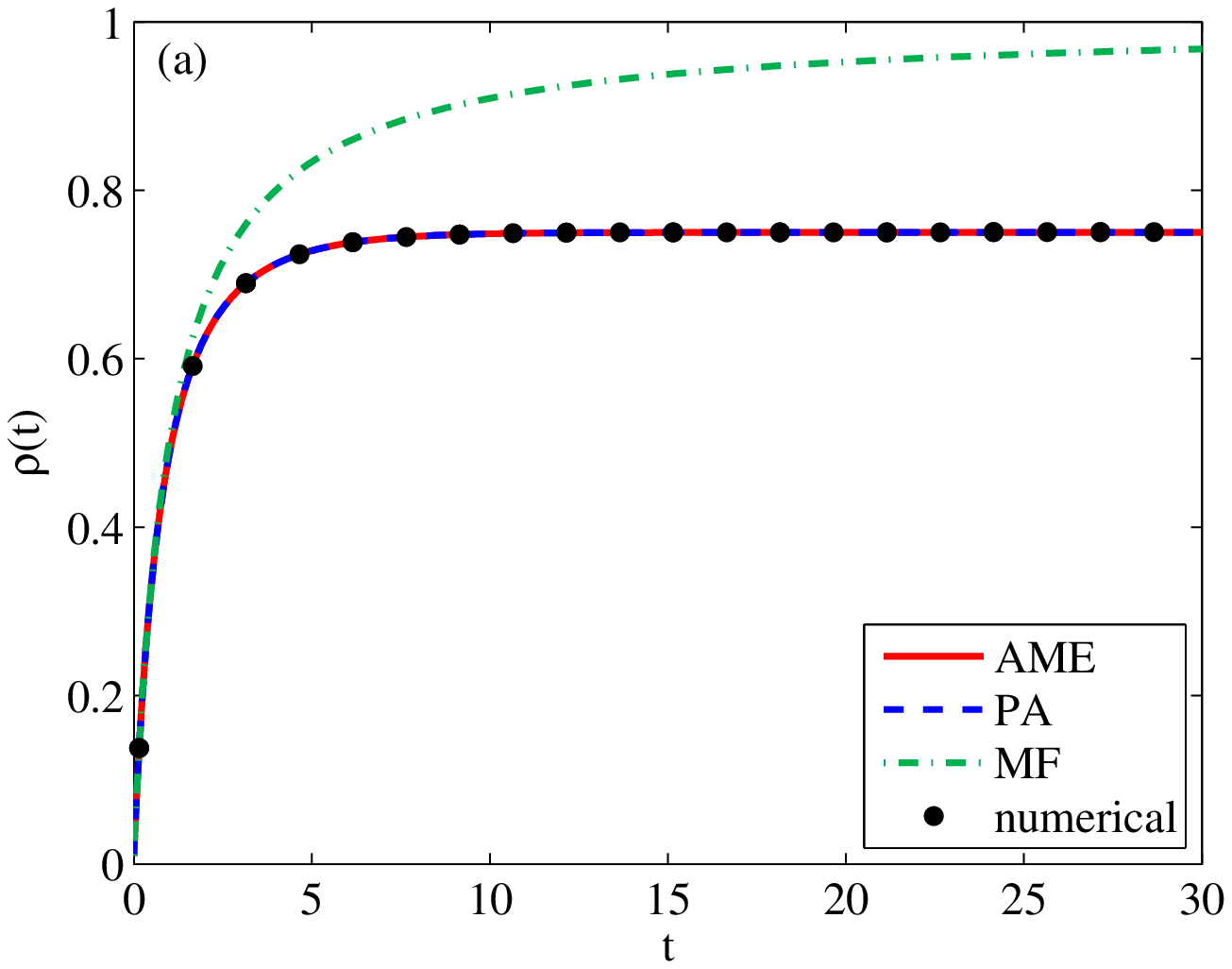,width=8cm}
\epsfig{figure=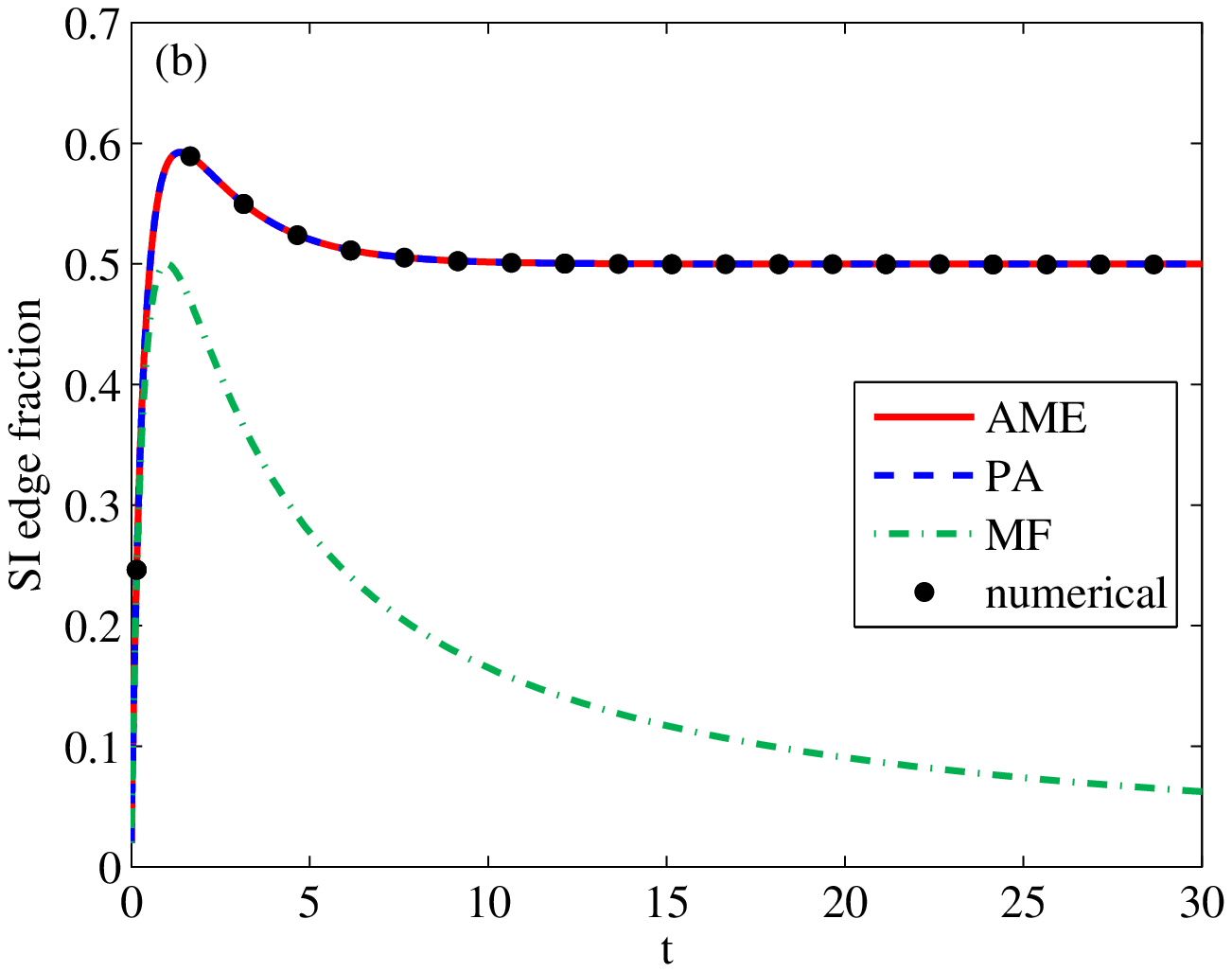,width=8cm}
\caption{ Bass diffusion dynamics on a 4-regular random graph (or Bethe lattice, $P_k=\delta_{k,4}$), with parameters $c=1$ and $d=-1/4$, and initial condition $\rho(0)=10^{-2}$. The AME and PA solutions for the fraction of adopted nodes are both given by the explicit formula (\ref{11}).}\label{fig:Bass}
\end{figure}

The parameters $d_k$ in (\ref{lin}) may be negative (provided that all $F_{k,m}$ values are non-negative). As an example, consider the Bass diffusion model on a $z$-regular random graph with $c=1$ and $d=-1/z$. This might model, for example, the adoption by indie music fans of a new band, where the attractiveness of the band decreases as the number of neighbors $m$ who have already ``jumped on the bandwagon'' increases; see \cite{Dodds12} for another approach to this aspect of social contagion modelling, which they call ``limited imitation contagion''. The expected fraction of fans $\rho(t)$ is given explicitly by Eq.~(\ref{eq:6}); note for these parameters the entire network does not become infected, see Figure~\ref{fig:Bass}.

\begin{figure}
\centering
\epsfig{figure=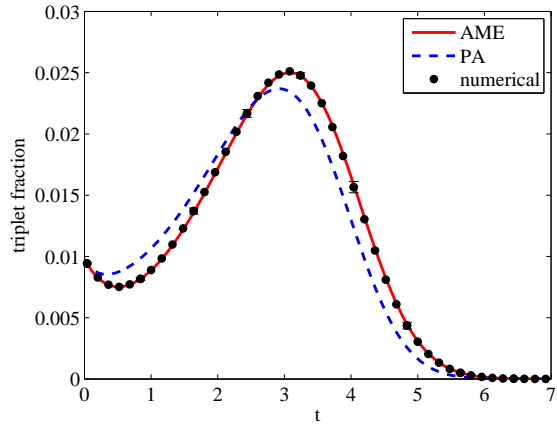,width=8cm}
\caption{Fraction of triples that are of \emph{S-I-S} type, for SI dynamics on a 3-regular random graph. }\label{fig_triples_ex}
\end{figure}

It is interesting to note that while, in the monotone case (\ref{lin}), the AME solutions for $s_{k,m}(t)$ are exactly reproduced by the PA equations, the corresponding $i_{k,m}(t)$ variables are not necessarily equal to their respective PA ansatz $\rho_k(t) B_{k,m}(q_k(t))$. Focussing on the SI model, the effect of the non-binomial $m$ distribution in $i_{k,m}$ is not visible in Figure~\ref{fig:SI}, as the quantities shown there can be written in terms of $s_{k,m}$ only; indeed, it is necessary to examine connected triples of nodes to demonstrate this effect \cite{Taylor12b}. Figure~\ref{fig_triples_ex} shows the fraction of connected triples (defined by choosing a node at random, then randomly choosing two of its neighbors) that are of type \emph{S-I-S}: the chosen (middle) node being infected, while both chosen neighbors are susceptible. In the AME formulation, this fraction is given by
$2 \sum_k P_k \sum_{m=0}^k (k-m)(k-m-1) i_{k,m}/\left<k(k-1)\right>$, whereas the corresponding PA version is $2 \sum_k P_k \rho_k k(k-1) (1-q_k)^2/\left<k(k-1)\right>$. The differences seen in Fig.~\ref{fig_triples_ex}---and the close fit of AME results to numerical simulations---indicate that the match between AME and PA results in Figs.~\ref{fig:SI} and \ref{fig:Bass} is due to the binomial $m$ distribution $s_{k,m}$ for susceptible nodes, but the corresponding distribution $i_{k,m}$ for infected nodes is \emph{not} binomial in $m$.

\section{General dynamics and equilibrium spin models} \label{sec:general}
We consider in this section the more general  case of non-monotone dynamics, where none of the rates $F_{k,m}$ and $R_{k,m}$ are identically zero. Inserting the PA ansatz for $s_{k,m}$ and $i_{k,m}$ into the AME equations (\ref{seqns})--(\ref{ieqns}) we find, after dividing by $(1-\rho_k)B_{k,m}(p_k)$, the condition
\begin{equation}
-F_{k,m} + R_{k,m}\frac{\rho_k}{1-\rho_k} \frac{B_{k,m}(q_k)}{B_{k,m}(p_k)} = c_k^{(1)} + c_k^{(2)} m \label{G1},
\end{equation}
where $c_k^{(1)}$ and $c_k^{(2)}$ represent combinations of terms that depend on $t$ and $k$, but are independent of $m$. Similarly, using the PA ansatz in the AME equation for $i_{k,m}$ yields
\begin{equation}
F_{k,m} \frac{1-\rho_k}{\rho_k} \frac{B_{k,m}(p_k)}{B_{k,m}(q_k)} -R_{k,m}= c_k^{(3)} + c_k^{(4)} m \label{G2},
\end{equation}
where, like Eq.~(\ref{G1}), the left-hand side is an exponential function of $m$ (because of the presence of the $B_{k,m}$ functions, see Eq.~(\ref{binomialB})), while the right-hand side is linear in $m$. Bearing in  mind that the transition rates $F_{k,m}$ and $R_{k,m}$ are time-independent, it is not generally possible to simultaneously solve equations (\ref{G1}) and (\ref{G2}) to obtain constant transition rates. We conclude that in this general case, AME solutions are not exactly equal, for all times, to the corresponding PA solutions.

However, an important special case occurs if we restrict our attention to the steady-state limit $t\to \infty$. The AME solutions and PA solutions can be identical in the limit $t \to \infty$ (despite being different at finite $t$) if equations (\ref{G1}) and (\ref{G2}) are simultaneously satisfied in the steady state. The $m$ dependence of the left- and right-hand sides then requires that
\begin{equation}
\frac{F_{k,m}}{R_{k,m}} = \frac{\overline{\rho}_k}{1-\overline{\rho}_k}\frac{B_{k,m}(\overline{q}_k)}{B_{k,m}(\overline{p}_k)},
\label{C1}
\end{equation}
and
$\overline{c}_k^{(1)}=\overline{c}_k^{(2)}=\overline{c}_k^{(3)}=\overline{c}_k^{(4)}=0$,
where the overbar denotes the steady-state limit of the corresponding variable, e.g., $\overline{p}_k = \lim_{t\to \infty} p_k(t)$.

The conditions $\overline{c}_k^{(1)}=\overline{c}_k^{(2)}=0$ can be shown to imply that
$ \overline{p}_k/(1-\overline{p}_k) = \overline{\beta}^s/\overline{\gamma}^s$, and the $k$-independence of the right-hand side implies that $\overline{p}_k$ must, in fact, be independent of $k$:
\begin{equation}
\overline{p}_k = \overline{p}\quad \text{ for all }k. \label{pindepk}
\end{equation}
Similarly, an analysis of the conditions $\overline{c}_k^{(3)}=\overline{c}_k^{(4)}=0$ shows that the steady-state values of $q_k$ must also be independent of $k$, i.e.,  $\overline{q}_k = \overline{q}$ for all $k$.

\begin{figure}
\centering
\epsfig{figure=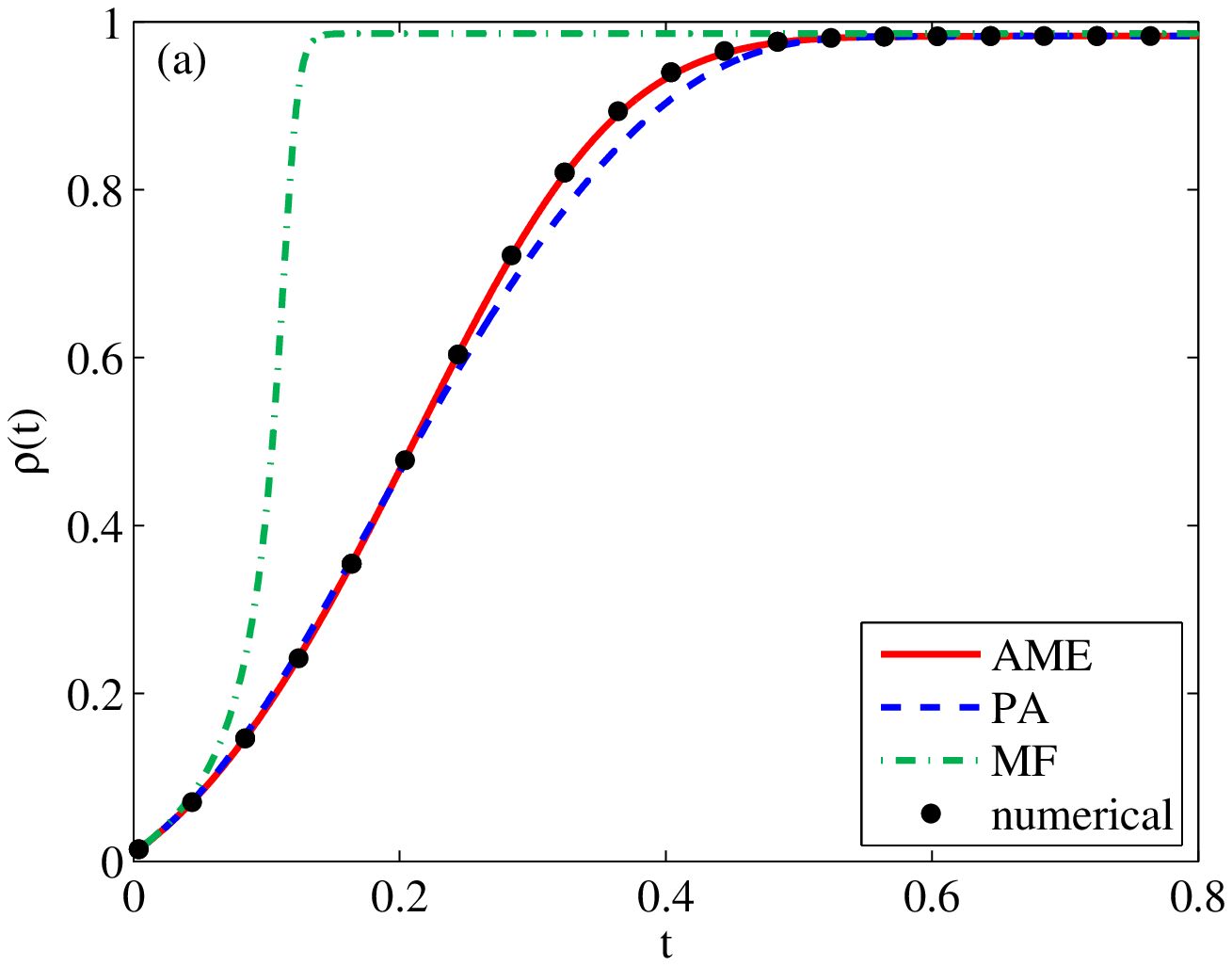,width=8cm}
\epsfig{figure=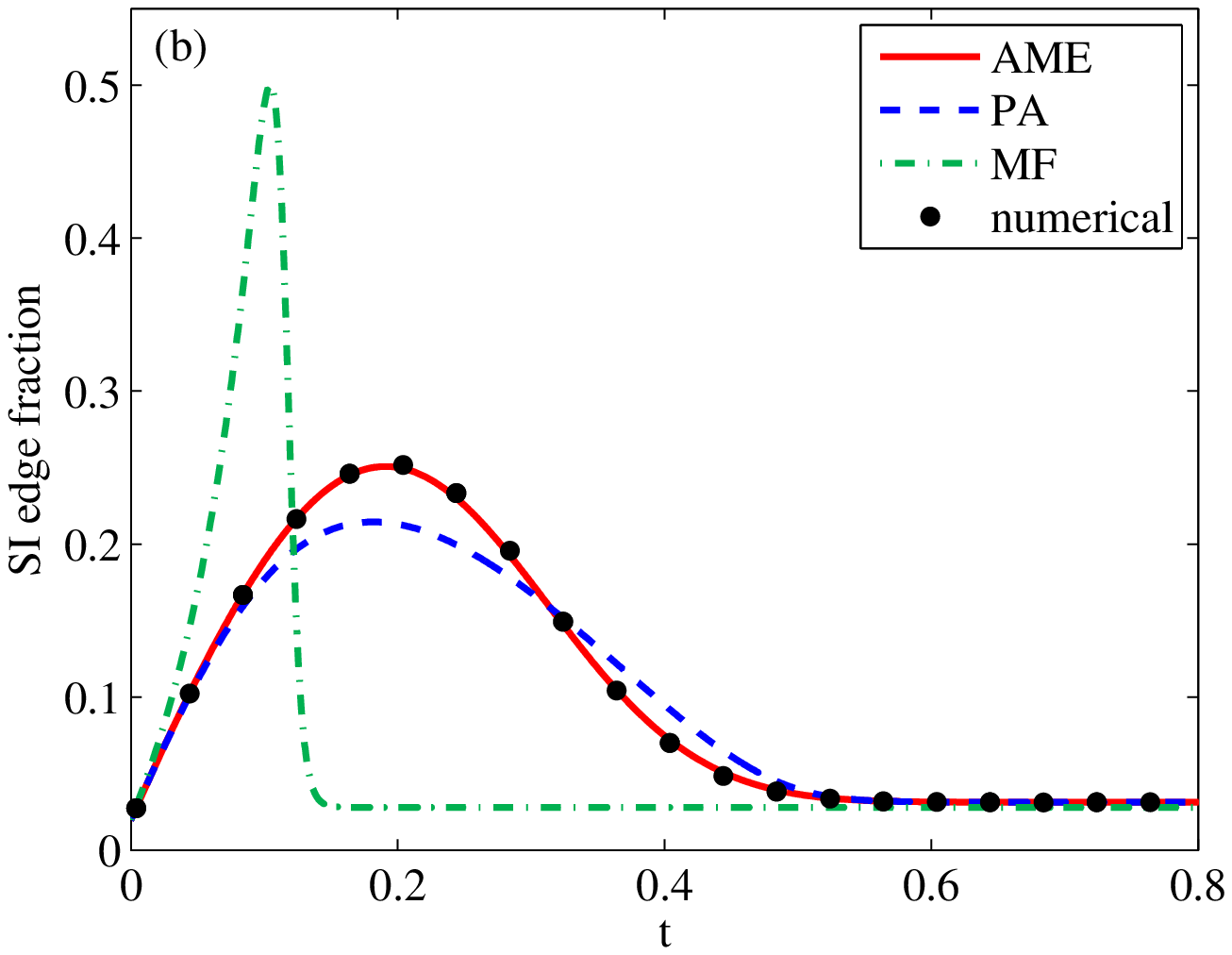,width=8cm}
\caption{ Modified SIS-type model described by the rates (\ref{FRgenSI}), on a 3-regular random graph. The parameters are $\lambda=6$ and $\mu=3$; initial state has $\rho(0)=10^{-2}$.}\label{fig:genSI}
\end{figure}

Replacing $\overline{p}_k$ and $\overline{q}_k$ in Eq.~(\ref{C1}) with these simplifications, we can rewrite the condition on the rates as
\begin{equation}
\frac{F_{k,m}}{R_{k,m}} = b_k a^m , \label{T}
\end{equation}
where
\begin{equation}
b_k= \frac{\overline{\rho}_k}{1-\overline{\rho}_k} \left( \frac{1-\overline{q}}{1-\overline{p}}\right)^k \label{beqn}
\end{equation}
and
\begin{equation}
a=\frac{\overline{q}(1-\overline{p})}{\overline{p}(1-\overline{q})} \label{aeqn}.
\end{equation}
It is demonstrated in Appendix~\ref{appB} that Eq.~(\ref{T}) is precisely the condition for microscopic reversibility (detailed balance) of the stochastic dynamics, i.e., for the dynamics to correspond to an equilibrium spin-flip model. Moreover, we show in Appendix~\ref{appC} that the steady solution of the AME for transition rates obeying condition (\ref{T}) can be  fully specified in a simple manner, by the equation
\begin{equation}
\overline{\rho}_k = \frac{b_k}{b_k + \left(1-\overline p+\overline p a \right)^{-k}}, \label{rhoksoln}
\end{equation}
where $\overline p$ is a solution of the algebraic equation
\begin{equation}
\frac{\overline p(1-\overline p+\overline p a)}{1-{\overline p}^2+{\overline p}^2 a} = \sum_k \frac{k}{z}P_k \frac{b_k}{b_k+(1-{\overline p}+{\overline p}a)^{-k}} .\label{wsoln}
\end{equation}
As an example of dynamics obeying condition (\ref{T}), we introduce a modified model of susceptible-infected-susceptible type, with
\begin{equation}
F_{k,m}=\lambda^m \quad\text{ and }\quad R_{k,m} = \mu \label{FRgenSI}
\end{equation}
for positive constants $\lambda$ and $\mu$. Like the standard SIS disease-spread model (see Table~\ref{tabFR}), recovery of infected nodes occurs at a constant rate $\mu$. However, in constrast to the linear-in-$m$ dependence of the SIS infection rate $F_{k,m}$, here the probability of infection is assumed to grow exponentially with $m$. While admittedly artificial, such dynamics might find application in fitting data from social contagion experiments on networks, where the dependence of the rates of $m$ is not yet clear \cite{Romero11,Hodas12,Centola10}. Figure~\ref{fig:genSI} shows that  the AME solutions and PA solutions indeed agree as $t\to \infty$, despite being different for finite $t$.

\begin{figure}
\centering
\epsfig{figure=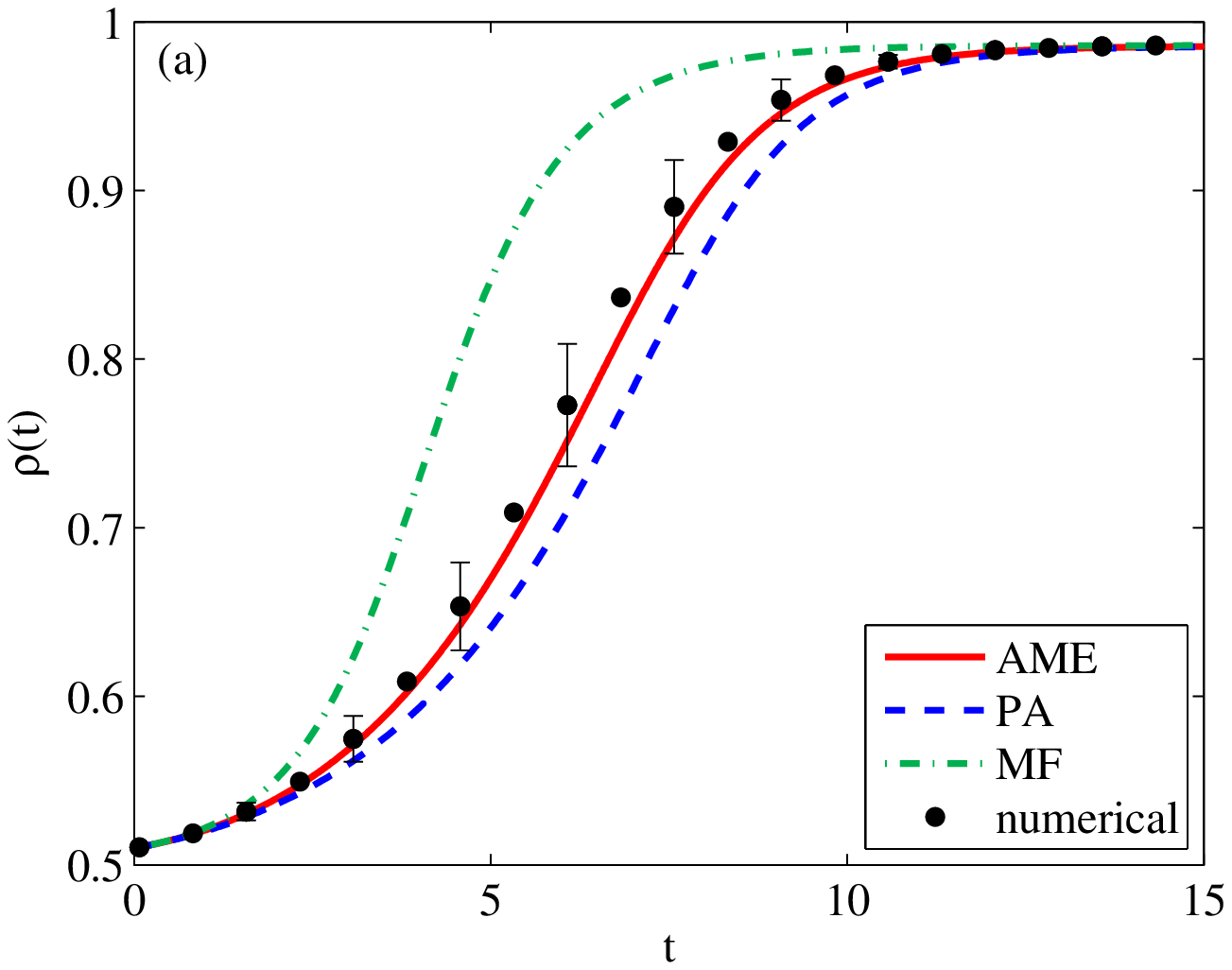,width=8cm}
\epsfig{figure=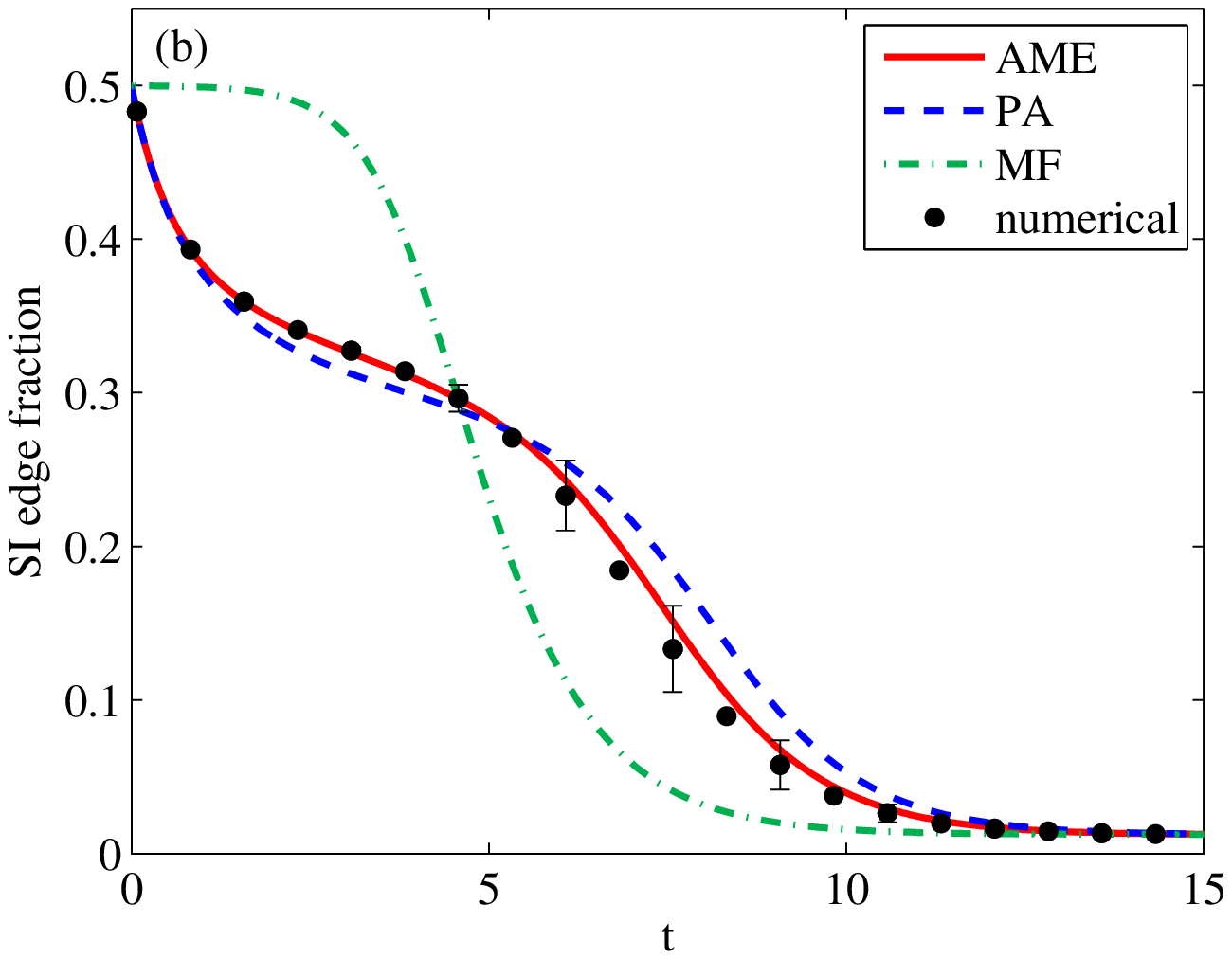,width=8cm}
\caption{ Glauber dynamics for the Ising spin model dynamics on a Poisson (Erd\"{o}s-R\'{e}nyi) random graph with mean degree $z=7$. The interaction parameter $J$ is set to 1,  the temperature $T$ is $2/\log(2.5)\approx 2.18$, and the initial fraction of up-spins is $0.51$.
}\label{fig:Ising}
\end{figure}

Other important examples of equilibrium models obeying (\ref{T}) are given by the Glauber and Metropolis dynamics for the Ising spin model (see Table~\ref{tabFR} and Fig.~\ref{fig:Ising}), which both have $a=e^{\frac{4 J}{T}}$ and $b_k = e^{-\frac{2 J}{T}k}$ in Eq.~(\ref{T}). These are examples of spin models with up-down symmetry, and are considered in greater detail in Sec.~\ref{sec:spin} below.

\section{Models with up-down symmetry} \label{sec:spin}
Models with up-down symmetry have dynamics that are invariant to the swap of state labels (susceptible to infected, and vice versa) for all nodes. In \cite{Castellano12b,Vazquez08b} this symmetry is called ``$Z_2$-symmetry''; it is characteristic of the voter model and other opinion dynamics models; see the fourth column of Table~\ref{tabFR}. In terms of the transition rates, the symmetry condition implies that
\begin{equation}
R_{k,m} = F_{k,k-m} \quad \text{ for } m=0,\ldots,k \text{ and for all }k. \label{T2}
\end{equation}
Note that for such dynamics, the AME system (\ref{seqns})--(\ref{ieqns}) is invariant under the change of variables $s_{k,m} \mapsto i_{k,k-m}$ and $i_{k,m} \mapsto s_{k,k-m}$. Since the expected fraction of degree-$k$ nodes can be written as $\rho_k = \sum_{m=0}^k i_{k,m}= 1-\sum_{m=0}^k s_{k,m}$, this symmetry condition implies that a solution exists  of the AME with $\rho_k(t)=1/2$ for all $t$. However, this may not be the only possible solution: depending on the initial condition $\rho(0)$, and on the parameter regime, other solutions of the AME may also be found.

Focussing first on equilibrium spin models with up-down symmetry, which obey condition (\ref{T}) in addition to (\ref{T2}), we investigate the stability of the symmetric solution with $\overline \rho_k = 1/2$. First, note that there is only a single parameter in these models: putting $m=k/2$ for even $k$, or, for odd $k$, $m=(k-1)/2$ and then $m=(k+1)/2$, into Eq.~(\ref{T}) and imposing condition (\ref{T2}) immediately yields the necessary relation
\begin{equation}
b_k = a^{-\frac{k}{2}} \label{barel}
\end{equation}
between the parameters of equilibrium spin models. Using the steady-state solution of Sec.~\ref{sec:general}, it is possible to show that a critical value $a_c$ of the parameter $a$ exists: in the language of dynamical systems, this is a (pitchfork) bifurcation point \cite{Strogatzbook}. For parameter values $a$ with $a<a_c$, the symmetric solution $\rho = 1/2$ is stable, meaning that if $\rho(0)$ is close to 1/2, the steady-state solution will be $\overline \rho=1/2$. In spin models (where the magnetization can be written as $M=|2\rho-1|$) this regime is the paramagnetic (disordered) phase; for opinion models  in this regime  the two opinions coexist equally on the network. However, if the parameter $a$ exceeds the critical value $a_c$, then the  symmetric solution $\rho=1/2$ is unstable, and two other stable solutions, symmetric about $\rho=1/2$, exist. This is the ferromagnetic (ordered) phase; for opinion models, one of the two opinions dominates the other. The critical value $a_c$ gives the phase transition point, and the behavior of $\overline \rho$ near $a_c$ can be determined from Eq.~(\ref{wsoln}) (see Appendix~\ref{appD} for details) in a very similar fashion to the analysis of the Ising model in \cite{Leone02,Dorogovtsev02}, see also \cite{Dorogovtsev08}. The results of such analysis (see Appendix~\ref{appD}) may be summarized as follows. If the degree distribution $P_k$ possesses a finite fourth moment $\left<k^4\right> = \sum_k k^4 P_k$, then the phase transition is of mean-field type, with critical parameter
\begin{equation}
a_c = \left(\frac{\left< k^2\right>}{\left<k^2\right> - 2\left< k\right>}\right)^2 \label{acrit}
\end{equation}
and with $\overline \rho -1/2 \sim \pm (a-a_c)^{\frac{1}{2}}$ as $a \to a_c$ from above. Following \cite{Leone02,Dorogovtsev02} (see Appendix~\ref{appD}), if the network has a scale-free degree distribution $P_k \sim k^{-\gamma}$ as $k\to \infty$, then for exponents $ \gamma$ in the range $2<\gamma<3$, the critical point is $a_c=1$, with
$\overline \rho - 1/2\sim \pm (a-a_c)^\frac{1}{3-\gamma}$, while for exponents with $3<\gamma<5$, we have $a_c$ given again by Eq.~(\ref{acrit}), but with near-criticality scaling of $\overline \rho-1/2 \sim \pm (a-a_c)^\frac{1}{\gamma-3}$ as $a\to a_c$. The case $\gamma=3$ shows an infinite order transition at $a=1$, as discussed in \cite{Dorogovtsev02}.

As mentioned at the end of Sec.~\ref{sec:general}, the Ising model (Glauber or Metropolis dynamics) is of type (\ref{T2}), with temperature $T$ related to the parameter $a$ via $T=4 J /\ln a$, so $a=1$ corresponds to infinite temperature.  A social influence model of this type might be given, for example, by transition rates with the properties
\begin{equation}
R_{k,m} = F_{k,k-m} = F_{k,m}\quad\text{ for } m=0,\ldots,k \text{ and all }k. \label{a1ex}
\end{equation}
Here all $F_{k,m}$ values for $m=0$ to $\lfloor \frac{k}{2} \rfloor$   are free parameters; for example, the rates $F_{k,m}$ and $R_{k,m}$ could be given by Fig.~\ref{fig:FRschematic}(c), which is motivated by the data analysis in Figure~4 of \cite{VerSteeg11}. In any model satisfying (\ref{a1ex}) the parameter $a$ of (\ref{T2}) is equal to 1, and by the results above the model is in  the paramagnetic (disordered) phase for all network topologies. However, Eq.~(\ref{acrit}) shows that $a=1$ is near the critical point of the system if the degrees in the network are very heterogeneous (so that $\left< k^2 \right> \gg \left<k\right>$), and indeed the model is poised precisely at criticality ($a_c=1$) on scale-free networks with infinite-variance degree distributions.

\begin{figure}
\centering
\epsfig{figure=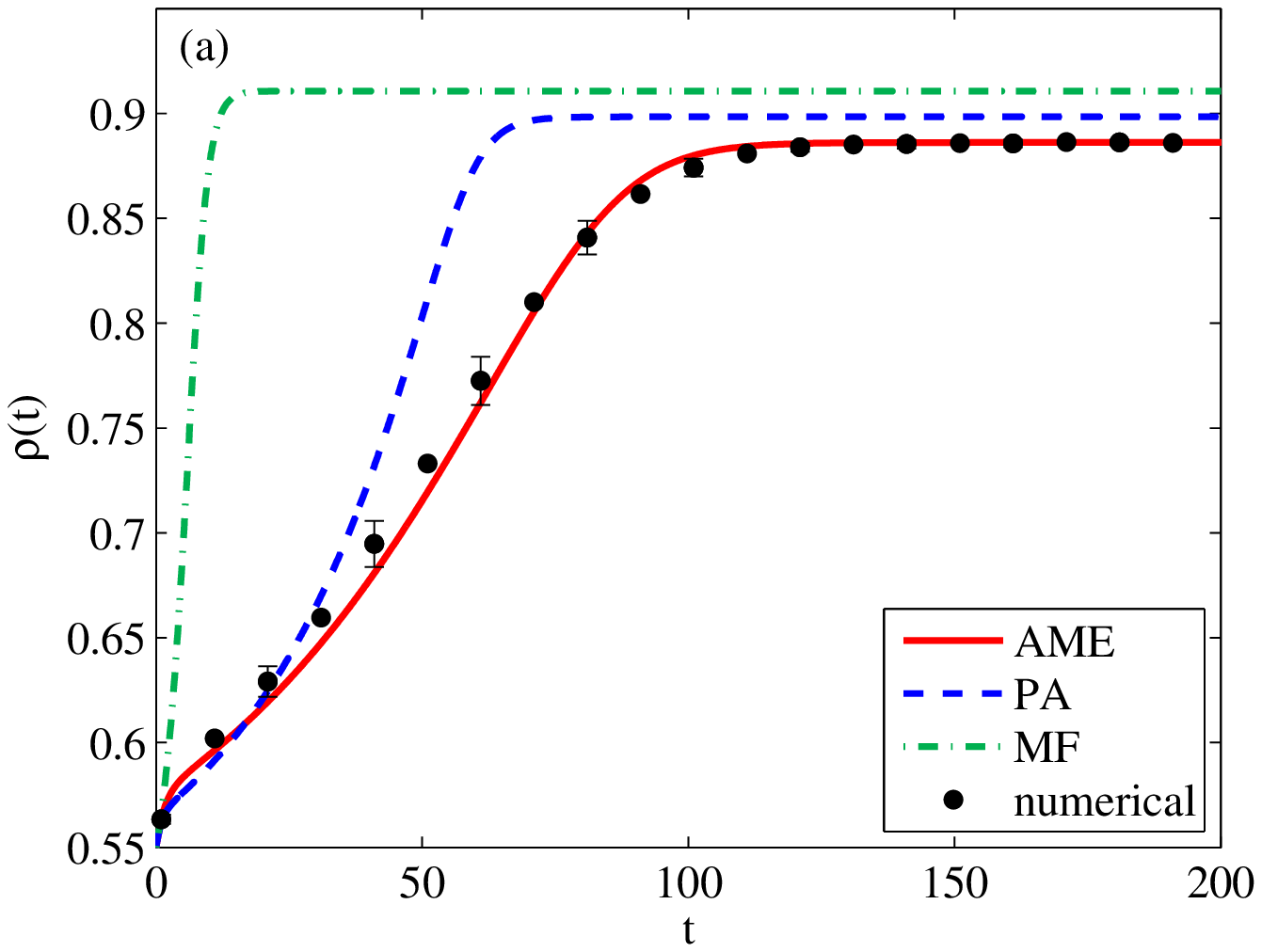,width=8cm}
\epsfig{figure=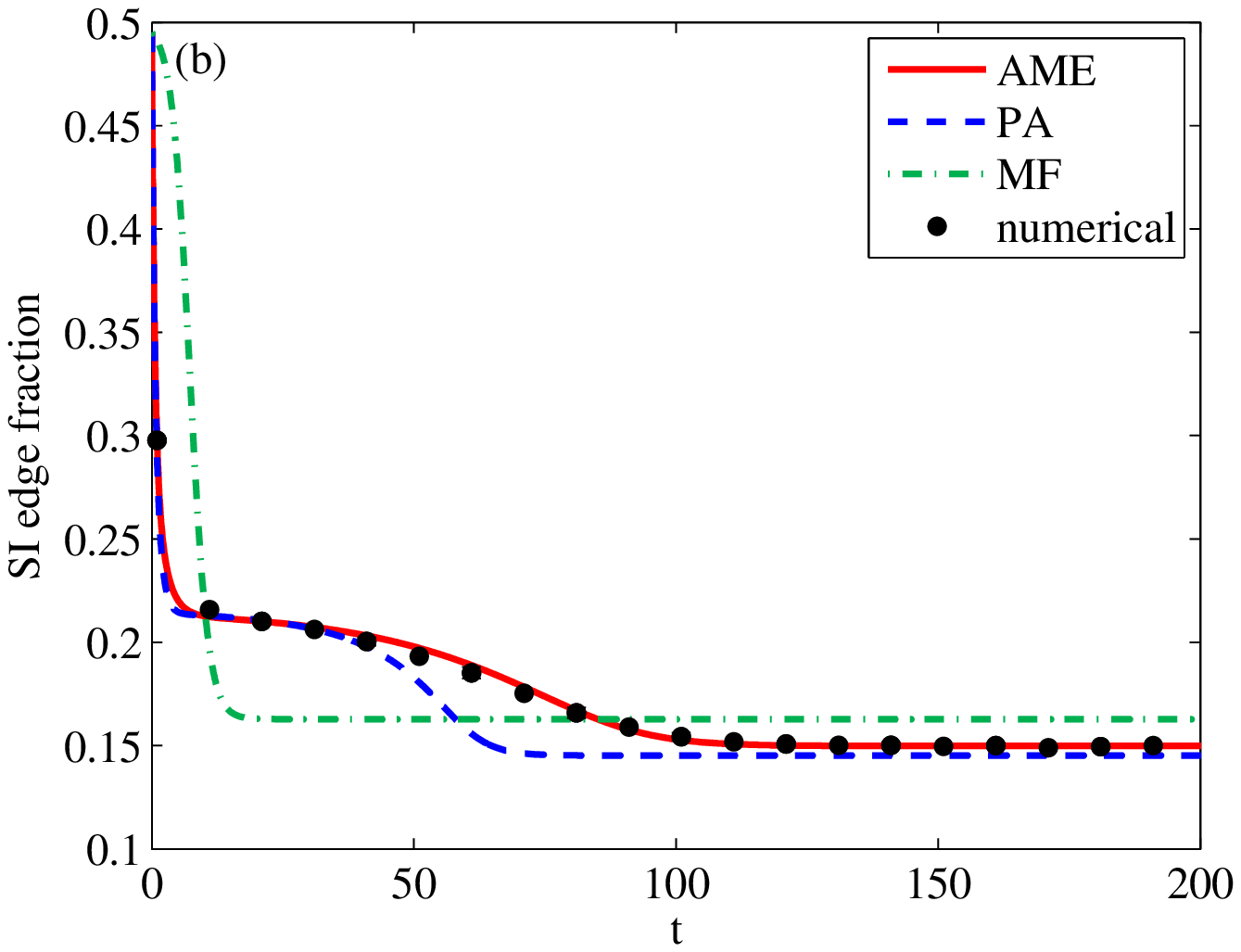,width=8cm}
\caption{ Majority-vote model on 3-regular random graph, with $\rho(0)=0.55$ and noise parameter $Q=0.07$. This non-equilibrium model does not obey condition (\ref{T2}) and the AME and PA solutions are not equal, even as $t\to \infty$, in contrast to Figs.~\ref{fig:genSI} and \ref{fig:Ising}.}\label{fig:maj_vote}
\end{figure}

For symmetric models that do not obey the equilibrium condition (\ref{T}),  the PA and AME solutions are different, even as $t\to \infty$, see Fig.~\ref{fig:maj_vote} for an example using majority-vote dynamics.
The solutions of the PA equation are therefore of limited usefulness, but it is nevertheless interesting that for $z$-regular graphs an expression for the critical point can be explicitly obtained. In Appendix~\ref{appE} we show that the PA critical point of symmetric models on such networks occurs precisely when the steady-state PA paramater $p$ has the value
\begin{equation}
\overline p_{c} =\frac{z-2}{2 z - 2} \label{z1},
\end{equation}
and the location of the critical point in parameter space is given by the solutions $F_{z,m}$ (for $m=0,\ldots,z$) of the implicit equation
\begin{equation}
\sum_{m=0}^z \left(1-\frac{2 m}{z}\right) F_{z,m} B_{z,m}\left(\frac{z-2}{2 z - 2}\right) = 0 \label{z2}.
\end{equation}
Using the infection rate $F_{k,m}$ for
the majority-vote model (see Table~\ref{tabFR}), for example, in Eq.~(\ref{z2}), gives an explicit expression for the PA
critical noise parameter $Q$:
\begin{equation}
Q_c = \left[ 1-\frac{\sum_{m=0}^{\lfloor \frac{z-1}{2}\rfloor}\left(1-\frac{2m}{z}\right) B_{z,m}\left(\frac{z-2}{2z-2}\right)}{\sum_{m=\lceil \frac{z+1}{2}\rceil}^{z}\left(1-\frac{2m}{z}\right) B_{z,m}\left(\frac{z-2}{2z-2}\right)} \right]^{-1}.
\end{equation}
Equation (\ref{z2}) can similarly be used to obtain analytical expressions for the PA critical points in other models on $z$-regular random graphs, e.g. Fig.~12 of \cite{Schweitzer09}, Fig.~1 of \cite{deOliveira93}, or Fig.~1 of \cite{Drouffe99}.

\section{Threshold dynamics} \label{sec:thresh}
Threshold models are often used to model propagation of fads or collective action through a population \cite{Granovetter78,Watts02,Centola07,Kleinbergbook,Galstyan07,Gleeson07}. Each node has a (frozen) threshold level; these thresholds may be chosen at random (e.g., from a Gaussian distribution), or assigned in some other way (perhaps depending on the degree of the node). In the \emph{asynchronous-updating} version of these models, a fraction $dt$ of nodes are randomly chosen for updating in each time step \cite{footnote5}. When chosen for updating, an inactive node becomes active (with probability 1) if $m$, the number of its neighbors who are active, exceeds the node's threshold \cite{footnote6}. Once activated, nodes cannot subsequently become inactive, so the dynamics are monotone, cf.~Sec.~\ref{sec:mono}. Unlike in Eq.~(\ref{lin}) of Sec.~\ref{sec:mono} however, the transition rate is not linear in $m$; in fact, it is given by
\begin{equation}
F_{\kv,m} = \left\{ \begin{array}{cc}
                    0 & \text{ if }m <M_\kv\\
                    1 & \text{ if }m\ge M_\kv,
                   \end{array} \right.
                   \label{2}
\end{equation}
to reflect the deterministic activation of a node (once it is chosen for updating) when $m$ exceeds the threshold level $M_\mathbf{k}$. We have introduced here the vector $\mathbf{k}$ to encode two properties defining a class of node: their degree $k$ (a scalar), and their \emph{type} $r$, which together determine the threshold $M_\mathbf{k}$ for such nodes. The types are assumed to be from a discrete set of possibilities: all nodes of type $r=1$, for example, might have the same threshold $M_1$, with all nodes of type $r=2$ having a common threshold $M_2$, with $M_2\neq M_1$. In this way, the set of all nodes may be partitioned into disjoint sets labelled by their degree and their type; in mathematical notation we combine these into a 2-vector, defining $\mathbf{k}=\{ k, r\}$ for the $\mathbf{k}$-class of nodes. All nodes in the same $\mathbf{k}$-class have the same degree and the same type, and therefore all share the same threshold $M_\mathbf{k}$.
 We generalize the degree distribution $P_k$ to the distribution $P_\mathbf{k}$, which gives the probability that a randomly-chosen node has vector $\mathbf{k}$ (i.e., has degree $k$ and type $r$).
 For example, if the thresholds of the nodes are randomly chosen, independent of their degrees, then the $P_\mathbf{k}$ distribution can be written as $P_\mathbf{k}=P_k P_r$, where $P_k$ is the degree distribution and $P_r$ is the probability that a node is of type $r$. By taking the discrete set of types to be sufficiently large, it is possible to approximate a continuous distribution of types or thresholds with a desired level of accuracy. With this extended notation, the AME approach can be generalized in an obvious manner, essentially replacing the scalar degree $k$ by the vector $\mathbf{k}$ as appropriate in Eq.~(\ref{eq:4}), to yield
\begin{equation}
\frac{d}{dt}{s}_{\kv,m} = -F_{\kv,m} s_{\kv,m} - \beta^s (k-m) s_{\kv,m} + \beta^s (k-m+1) s_{\kv,m-1} \quad \text{ for } m=0,\ldots,k, \label{1}
\end{equation}
with the rate $\beta^s$  given by
$
\beta^s = \frac{\sum_\kv P_\kv \sum_{m=0}^k (k-m) F_{\kv,m} s_{\kv,m}}{\sum_\kv P_\kv \sum_{m=0}^k (k-m) s_{\kv,m}}.
$
Note the sums here are over all $\mathbf{k}$-classes, i.e., over all degrees $k$ and all types $r$: $\sum_\mathbf{k}:=\sum_k \sum_r$.

In \cite{Gleeson08} (see also \cite{Hackett_thesis11}) it is argued that for no-recovery threshold models of the type (\ref{2})--(\ref{1}) described above, the fraction $\rho(t)$ of active nodes at times $t$ can be found by solving just two differential equations:
\begin{eqnarray}
\frac{d}{dt}\rho &=& h(\phi)-\rho \nonumber\\
\frac{d}{dt} \phi &=& g(\phi) - \phi , \quad \text{ with } \quad \phi(0)=\rho(0) = \sum_\kv P_\kv \rho_\kv(0), \label{3}
\end{eqnarray}
where
\begin{equation}
h(\phi) = \sum_\kv P_\kv \left[ \rho_\kv(0) + (1-\rho_\kv(0)) \sum_{m\ge M_\kv} B_{k,m}(\phi) \right] \label{4}
\end{equation}
and
\begin{equation}
g(\phi) = \sum_\kv \frac{k}{z} P_\kv \left[ \rho_\kv(0)+(1-\rho_\kv(0)) \sum_{m\ge M_\kv} B_{k-1,m}(\phi) \right] \label{5}.
\end{equation}
Here $\rho_\kv(0)$ is the fraction of nodes with vector $\mathbf{k}$ that are activated (infected) at time $t=0$; as in \cite{Hackett_thesis11}, we generalize the usual infected fraction $\rho(0)$---that is used elsewhere in this paper---to allow for possible dependence on the degree or type of the nodes chosen to ``seed'' the contagion.

\begin{figure}
\centering
\epsfig{figure=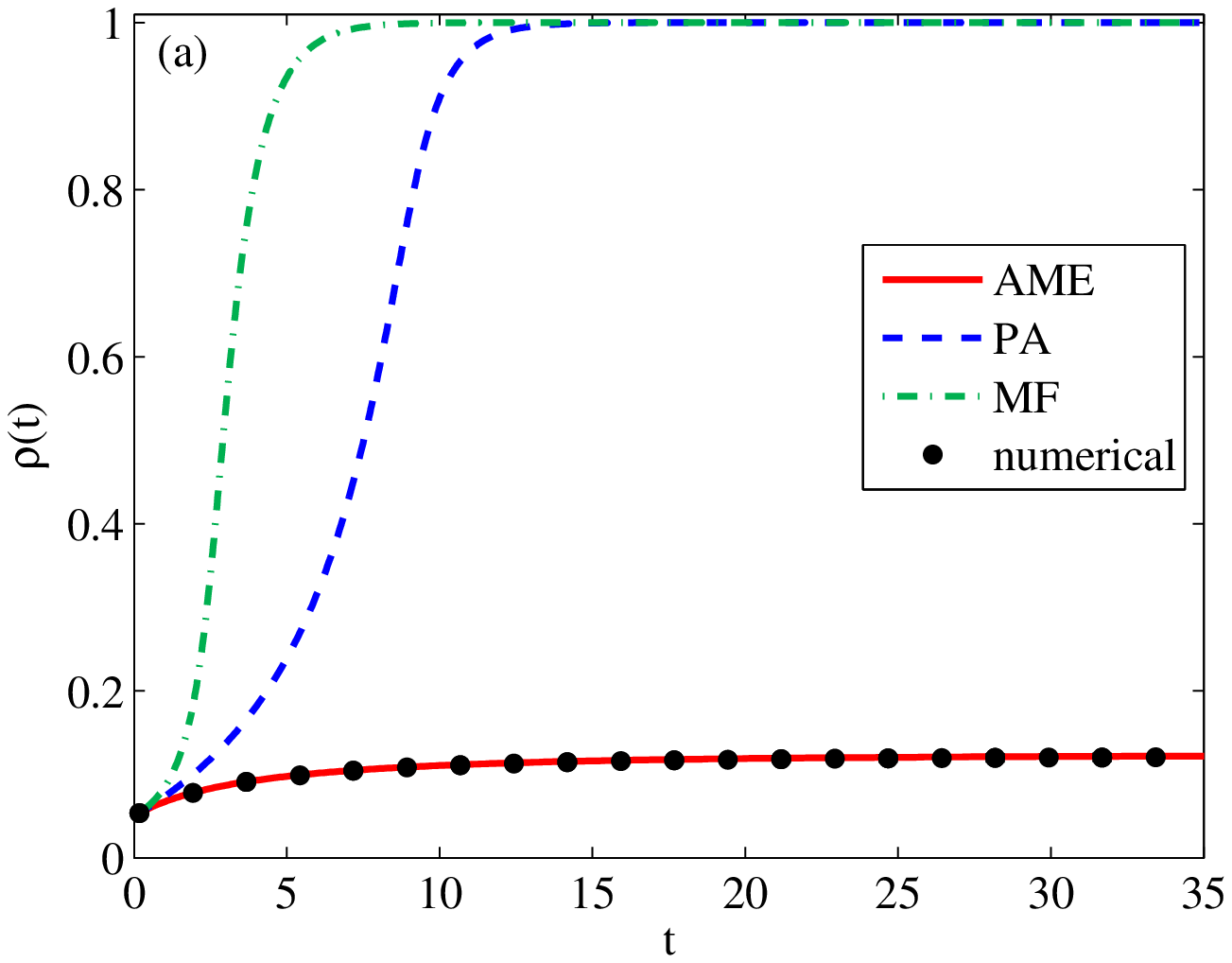,width=8cm}
\epsfig{figure=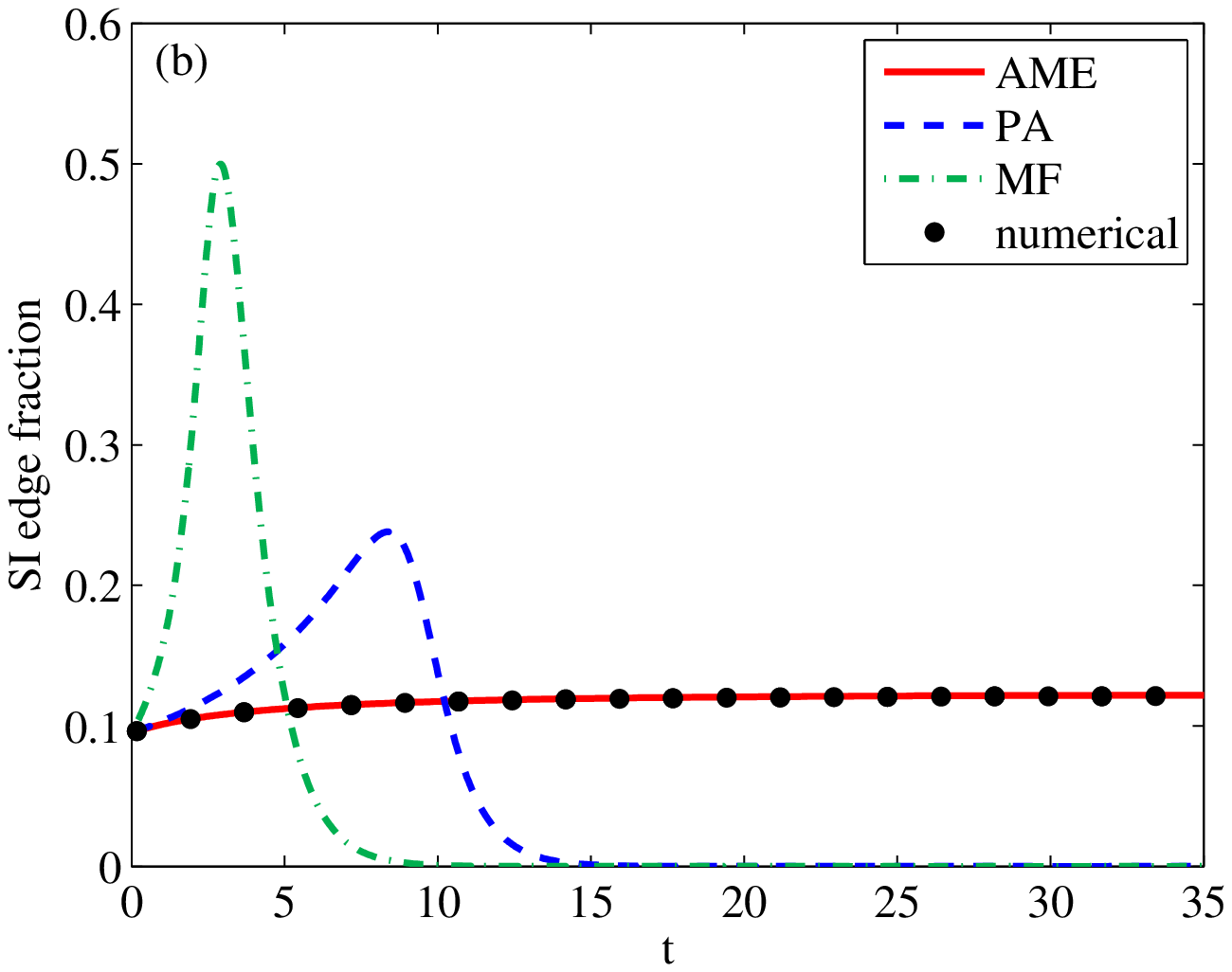,width=8cm}
\caption{ Threshold model on a 5-regular random graph, with all nodes having the same threshold $M=2$; the initial condition is $\rho(0)=0.05$. The AME solution is identical to the solution of the 2-dimensional system (\ref{3}).}\label{fig:thresh}
\end{figure}

In Appendix~\ref{appF} we demonstrate that Eqs.~(\ref{2}) and (\ref{1}) reduce to (\ref{3})--(\ref{5}) through an exact solution of (\ref{1}) given by
\begin{equation}
s_{\kv,m}(t) = \left( 1 - \rho_\kv(0)\right) B_{k,m} (\phi) \quad \text{ for }m < M_\kv \label{6}.
\end{equation}
The distribution of $s_{\kv,m}$ for $m \ge M_\kv$ is, in general, not of the binomial form (\ref{6}), but it can nevertheless be given explicitly as detailed in  Appendix~\ref{appF}. This means that the reduced-dimension system (\ref{3}) is not precisely of the PA type (\ref{PA}), but it enables efficient and very accurate solution of threshold-dynamics models, see the example in Fig.~\ref{fig:thresh}.

\section{Conclusions} \label{sec:conclusion}

We have pointed out that many stochastic binary-state dynamical systems on networks can be described using the transition rates $F_{k,m}$ and $R_{k,m}$ introduced in Sec.~\ref{sec:intro}. The main results of this paper are the identification of dynamics for which the full approximate master equation system (\ref{1}) can be reduced, without loss of accuracy, to a lower-dimensional system of equations, occasionally even yielding closed-form solutions (such as Eqs.~(\ref{eq:6}) and (\ref{z2})). We showed in Sec.~\ref{sec:mono} that the pair approximation system exactly matches the AME results for nodes in one state (e.g., the susceptible state) if the dynamics are monotone (i.e., the recovery rate $R_{k,m}$ is identically zero) and the infection rate is linear in the number of infected neighbors ($F_{k,m}=c_k + d_k m$). The classical SI model and the Bass diffusion model are of this type---see Figs.~\ref{fig:SI} and \ref{fig:Bass}---and for Bethe lattices ($z$-regular random graphs) Eq.~(\ref{eq:6}) explicitly gives the expected fraction of infected nodes. Interestingly, the PA solution is not exactly equal to the AME for nodes in the infected state, see Fig.~\ref{fig_triples_ex}.

In Sec.~\ref{sec:general} we showed that if both sets of transition rates $F_{k,m}$ and $R_{k,m}$ are non-zero, then the AME solutions cannot be given exactly by the PA ansatz for all time $t$. However, in the special case defined by Eq.~(\ref{T}), which corresponds to equilibrium stochastic dynamics, i.e., those obeying detailed balance (microscopic reversibility), the AME solutions reduce to the PA solution in the limit $t\to\infty$, see Figs.~\ref{fig:genSI} and \ref{fig:Ising}. This property allows us to perform bifurcation analysis for  systems with up-down ($Z_2$) symmetry (i.e., for dynamics obeying condition Eq.~(\ref{T2})), giving the explicit expression (\ref{acrit}) for the critical point where the steady-state limit of $\rho$ changes from the disordered (paramagnetic) state $\overline \rho=1/2$ to an ordered (ferromagnetic) state with $\overline \rho\neq 1/2$. The Glauber and Metropolis dynamics for the Ising spin model are important examples that obey both conditions (\ref{T}) and (\ref{T2}), and our results reproduce the critical Ising temperature that was found using very different methods (replica trick, recursion approach) in \cite{Leone02,Dorogovtsev02}. The analysis of \cite{Leone02,Dorogovtsev02} for the critical behavior on scale-free networks can also be applied here, and gives results described in Sec.~\ref{sec:spin} and Appendix~\ref{appD}. We highlight the fact that the critical point $a=1$ for networks with infinite-variance degree distributions is attainable in plausible social influence models of this type, as is the case $a<1$, although for the Ising model $a=1$ corresponds to infinite temperature and the regime $a<1$ is unphysical.  We also give an explicit expression (Eq.~(\ref{z2})) for the critical point (pitchfork bifurcation point) predicted by pair approximation of models with up-down symmetry---including non-equilibrium dynamics that obey (\ref{T2}) but not necessarily (\ref{T})---on $z$-regular random graphs, with the caveat that this PA critical point is not identical to the AME critical point except for the class of equilibrium models that obey  (\ref{T}) as well as (\ref{T2}).
Finally, in Sec.~\ref{sec:thresh}, we showed that the AME approach can be extended to include threshold models of adoption or fad diffusion, and that the AME system can be reduced to a system of only two differential equations, see Eqs.~(\ref{3}) and Fig.~\ref{fig:thresh}.

The exact agreement of AME and PA solutions---whether for all time as in Sec.~\ref{sec:mono}, or in the steady-state as in Sec.~\ref{sec:general}---implies that higher-order correlations (beyond nearest-neighbor) are correctly captured by PA in the cases we have identified. Indeed, the agreement of the theory curves with numerical simulation results in all these cases (for all time in  Figs.~\ref{fig:SI} and \ref{fig:Bass}, and as $t\to \infty$ in Figs.~\ref{fig:genSI} and \ref{fig:Ising}) is excellent. We interpret this as indicating that for the classes of dynamics where   PA is equal to AME, the results of PA are essentially exact. In contrast, in cases where PA and AME are not equal, as in Fig.~\ref{fig:maj_vote} for example, it is necessary to solve the full AME system to obtain high-accuracy approximations.

The present work is focussed only infinite, uncorrelated networks, with negligible levels of clustering (transitivity). Generalizing the AME approach to clustered network models and/or to networks with degree-degree correlations remains a challenge, as the added complexities will lead to even larger differential equation systems than needed for configuration model networks. Nevertheless, we hope the insights gained here may assist in generating and analyzing pair approximations for dynamics on such networks. Another direction for further work is to consider the effects of finite $N$ upon the various approximations used here. Although the local (node-level) dynamics are stochastic, the differential equations derived for the emergent dynamics are deterministic, because we assume the $N\to\infty$ limit. On smaller networks, stochastic effects will be important even at the global level and different approaches---such as branching processes  \cite{Noel12}---will be required to describe the variability of results across realizations.
\section*{Acknowledgements}

This work was partly funded by Science Foundation Ireland (awards 11/PI/1026 and 09/SRC/E1780) and  by the European Commission through FET-Proactive project PLEXMATH (FP7-ICT-2011-8; grant number
317614). We acknowledge the SFI/HEA Irish Centre for High-End Computing
(ICHEC) for the provision of computational facilities, and Sean Lyons, Davide Cellai, Sergey Melnik, Adam Hackett, and Peter Fennell for helpful discussions.

\appendix
\section{Derivation of Master equations} \label{appA}

\begin{figure}
\centering
\epsfig{figure=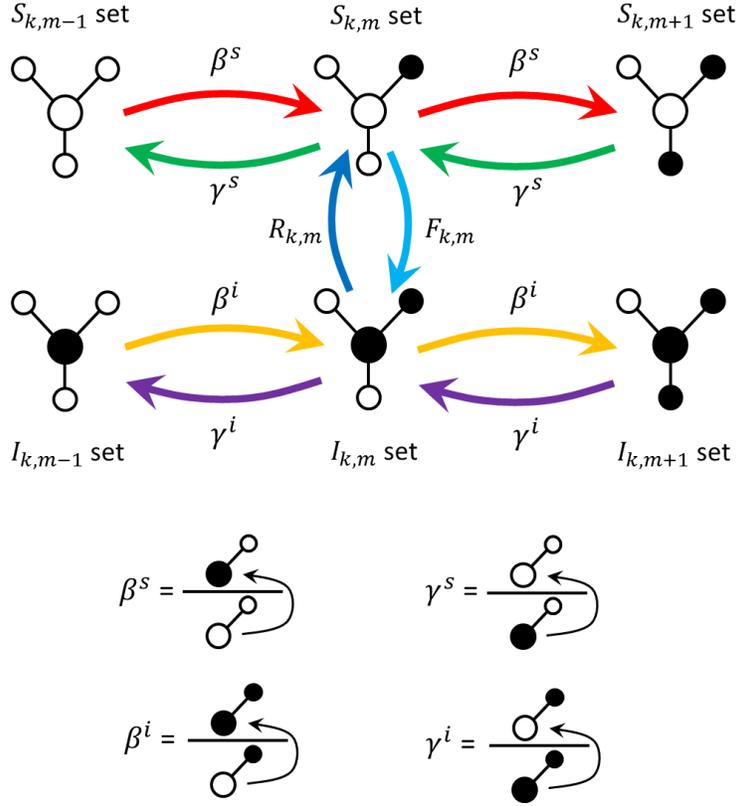,width=10cm}
\caption{Schematic of transitions to/from the $S_{k,m}$ and $I_{k,m}$ sets, as described in equations (\ref{Weqn}) through (\ref{betai}). For each set, the central (Ego) node is shown along with some of its neighbors: white nodes are susceptible/inactive/spin-down and black nodes are infected/active/spin-up. See also Fig.~1 of \cite{Marceau10}.}\label{fig_schematic}
\end{figure}

We consider binary-state dynamics  on static, undirected, connected networks in the limit of infinite network size (i.e., $N\to \infty$, where $N$ is the number of nodes in the network). For convenience, we call the two possible node states \emph{susceptible} and \emph{infected}, as is common in disease-spread models. However, this approach also applies to other binary-state dynamics, such as spin systems \cite{deOliveira93},  where each node may be in the $+1$ (spin-up) or the $-1$ (spin-down) state.
 The networks have degree distribution $P_k$ and are generated by the configuration model \cite{Bender78, Bollobas80,Newmanbook}. Dynamics are stochastic, and are defined by infection and recovery probabilities $F_{k,m}$ and $R_{k,m}$, which depend on the degree $k$ of a node, and on the current number $m$ of infected neighbors of the node, see Sec.~\ref{sec:derivation}.

We now proceed to derive the approximate master equations for dynamics of this type, closely following the approach used in \cite{Marceau10,Lindquist11} for SIS dynamics.
Let $S_{k,m}$ (resp. $I_{k,m}$) be the set of nodes which are susceptible (resp. infected), have degree $k$, and have $m$ infected neighbors. To quantify the size of these sets, define $s_{k,m}(t)$ (resp. $i_{k,m}(t)$) as the fraction of $k$-degree nodes that are susceptible (resp. infected) at time $t$ and have $m$ infected neighbors. Then the fraction $\rho_k(t)$ of $k$-degree nodes that are infected at time $t$ is given by
\begin{equation}
\rho_k(t) = \sum_{m=0}^k i_{k,m} = 1 - \sum_{m=0}^k s_{k,m}, \label{rhok}
\end{equation}
and the fraction of infected nodes in the whole network is found by summing over all $k$-classes:
\begin{equation}
\rho(t) = \sum_k P_k \, \rho_k(t).
\end{equation}
If a randomly-chosen fraction $\rho(0)$ of nodes are initially infected, then the initial conditions for $s_{k,m}$ and $i_{k,m}$ are easily seen to be
\begin{equation}
s_{k,m}(0)=\left( 1- \rho(0) \right) B_{k,m}(\rho(0)), \quad\quad i_{k,m}(0) = \rho(0) B_{k,m}(\rho(0)), \label{ICs}
\end{equation}
where  $B_{k,m}(q)$  is the binomial distribution introduced in Eq.~(\ref{binomialB}). Note that we can also calculate the number of edges of various types using this formalism. For example, the number of edges in the network which join a susceptible node to an infected node (we call these \emph{$S$-$I$ edges} for short) can be expressed in two equivalent ways:
\begin{equation}
N \sum_k P_k \sum_{m=0}^k m\, s_{k,m}\quad \text{ or } \quad N \sum_k P_k \sum_{m=0}^k (k-m) \,i_{k,m}. \label{star}
\end{equation}
The first of these expressions, for example, follows from noting that in a sufficiently large network that there are $N P_k$ nodes of degree $k$, of which a fraction $s_{k,m}$ are susceptible and have $m$ infected neighbors. Each such node contributes $m$ edges to the total number of $S$-$I$ edges. Similar expressions may also be given for the number of $S$-$S$ and $I$-$I$ edges in the network. We note that the equivalence of the two expressions in (\ref{star}) is preserved by the evolution equations described below.

Next, we examine how the size of the $S_{k,m}$ set changes in time. We write the general expression
\begin{eqnarray}
s_{k,m}(t+dt) &=& s_{k,m}(t) - W(S_{k,m}\!\to\! I_{k,m})\,s_{k,m}\,dt+W(I_{k,m}\!\to\! S_{k,m}) \, i_{k,m}\, dt \nonumber\\
& & \hspace{1cm}- W(S_{k,m}\!\to\! S_{k,m+1}) \, s_{k,m}\, dt + W(S_{k,m-1}\!\to\! S_{k,m})\, s_{k,m-1}\, dt \nonumber\\
& & \hspace{1cm}- W(S_{k,m}\!\to\! S_{k,m-1}) \, s_{k,m}\, dt + W(S_{k,m+1}\!\to\! S_{k,m})\, s_{k,m+1}\, dt \label{Weqn}
\end{eqnarray}
to reflect all the transitions whose rate is linear in $dt$ (all other state-transitions are negligible in the $dt \to 0$ limit), see Fig.~\ref{fig_schematic}. Here $W(S_{k,m}\to I_{k,m})\, dt$, for example, is the probability that a node in the $S_{k,m}$ set at time $t$ moves to the $I_{k,m}$ set by time $t+dt$. It is clear from the definitions above that
\begin{equation}
W(S_{k,m}\!\to\!I_{k,m}) = F_{k,m} \quad \text{ and }\quad W(I_{k,m}\!\to\!S_{k,m}) = R_{k,m}.
\end{equation}
A node moves from the $S_{k,m-1}$ set to the $S_{k,m}$ set if it remains susceptible, while one of its susceptible neighbors becomes infected. Note this means that an $S$-$S$ edge changes to an $S$-$I$ edge. If we suppose that $S$-$S$ edges change to $S$-$I$ edges at a (time-dependent) rate $\beta^s$, we can write \cite{footnote7} 
\begin{equation}
W(S_{k,m}\!\to\!S_{k,m+1})=\beta^s(k-m)\quad \text{ and }\quad W(S_{k,m-1}\!\to\!S_{k,m}) = \beta^s (k-m+1),
\end{equation}
since nodes in the $S_{k,m}$ set have $k-m$ susceptible neighbors, while those in the $S_{k,m-1}$ set have $k-m+1$ susceptible neighbors. To calculate $\beta^s$, we count the number of $S$-$S$ edges in the network at time $t$, and then count the number of edges which switch from being $S$-$S$ edges to $S$-$I$ edges in the time interval $dt$; the probability $\beta^s\, dt$ is given by taking the ratio of the latter to the former, i.e.,
\begin{equation}
\beta^s\, dt = \frac{\sum_k P_k \sum_{m=0}^k (k-m) F_{k,m} \,s_{k,m}\, dt}{\sum_k P_k \sum_{m=0}^k (k-m) s_{k,m}} . \label{betas}
\end{equation}
A similar approximation is used to define $\gamma^s$, the rate at which $S$-$I$ edges change to $S$-$S$ edges due to the recovery of an infected node:
\begin{equation}
\gamma^s =  \frac{\sum_k P_k \sum_{m=0}^k (k-m) R_{k,m} \,i_{k,m}}{\sum_k P_k \sum_{m=0}^k (k-m) i_{k,m}} , \label{gammas}
\end{equation}
and we then write
\begin{equation}
W(S_{k,m}\!\to\!S_{k,m-1})=\gamma^s m\quad \text{ and }\quad W(S_{k,m+1}\!\to\!S_{k,m}) = \gamma^s (m+1).
\end{equation}
Taking the limit $dt\to 0$ of equation (\ref{Weqn}) gives the master equation for the evolution of $s_{k,m}(t)$ (see Fig.~\ref{fig_schematic}):
\begin{equation}
\frac{d}{d t}s_{k,m} = -F_{k,m} s_{k,m} + R_{k,m} i_{k,m} - \beta^s (k-m) s_{k,m} + \beta^s(k-m+1) s_{k,m-1} - \gamma^s m s_{k,m} + \gamma^s(m+1) s_{k,m+1}, \label{seqnsApp}
\end{equation}
 where $m$ is in the range $0,\ldots,k$ for each $k$-class in the network (and adopting the convention $s_{k,-1}\equiv s_{k,k+1}\equiv 0)$.
Applying identical arguments, \emph{mutatis mutandis}, to the set $I_{k,m}$, we derive the corresponding system of equations for $i_{k,m}(t)$:
\begin{equation}
\frac{d}{d t}i_{k,m} = -R_{k,m} i_{k,m} + F_{k,m} s_{k,m} - \beta^i (k-m) i_{k,m} + \beta^i(k-m+1) i_{k,m-1} - \gamma^i m i_{k,m} + \gamma^i(m+1) i_{k,m+1}, \label{ieqnsApp}
\end{equation}
for $m=0,\ldots,k$ and for each $k$-class in the network, with time-dependent rates $\beta^i$ and $\gamma^i$ defined through $s_{k,m}$ and $i_{k,m}$ as
\begin{equation}
\beta^i = \frac{\sum_k P_k \sum_{m=0}^k m\, F_{k,m} \,s_{k,m}}{\sum_k P_k \sum_{m=0}^k m\, s_{k,m}}
\quad \text{ and } \quad \gamma^i =  \frac{\sum_k P_k \sum_{m=0}^k m\, R_{k,m} \,i_{k,m}}{\sum_k P_k \sum_{m=0}^k m\, i_{k,m}} .
\label{betai}
\end{equation}

The approximate master equations (\ref{seqnsApp}) and (\ref{ieqnsApp}), with the time-dependent rates $\beta^s$, $\gamma^s$, $\beta^i$ and $\gamma^i$ (defined as nonlinear functions of $s_{k,m}$ and $i_{k,m}$), form a closed system of deterministic equations which, along with initial conditions (\ref{ICs}), can be solved numerically using standard methods \cite{AME_solve}. Note that the evolution equations are completely prescribed by the functions $F_{k,m}$ and $R_{k,m}$, and so this method can be applied to any stochastic dynamical process defined by transition rates $F_{k,m}$ and $R_{k,m}$. For the SIS model, equations (\ref{seqnsApp}) and (\ref{ieqnsApp}) were derived in \cite{Marceau10} (see also \cite{Noel09} and \cite{Lindquist11}), with additional terms to study adaptive rewiring of the network.

\section{Microscopic reversibility in equilibrium models} \label{appB}

\begin{figure}
\centering
\epsfig{figure=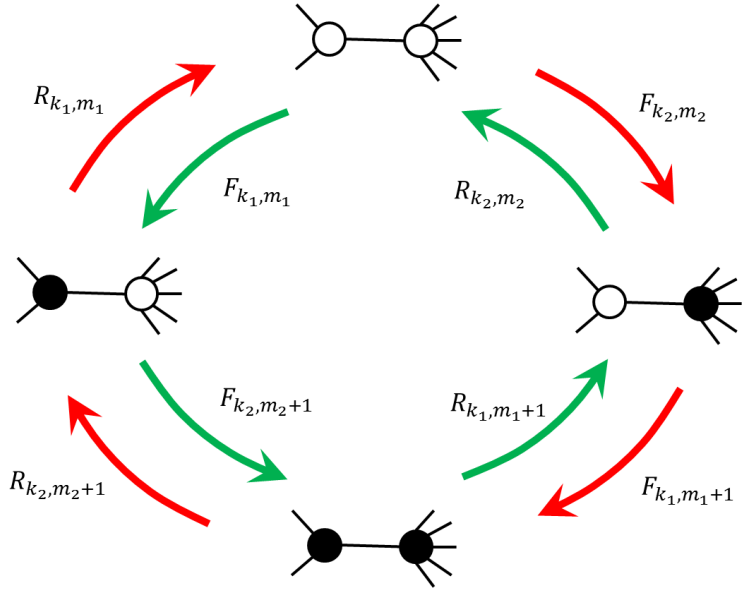,width=10cm}
\caption{Rates for state transitions of a connected pair of node, assuming that the states of all their other neighbors remain unchanged. Node 1 (left-hand node) has degree $k_1$ and has $m_1$ infected enighbors that are not shown. Node 2 (right-hand node) has degree $k_2$ and has $m_2$ neighbors---other than node 1---who are infected.}\label{fig2_inkscape}
\end{figure}
We consider transition rates $F_{k,m}$ and $R_{k,m}$ for which detailed balance holds, i.e., for which the dynamics exhibit microscopic reversibility \cite{Barratbook}. Figure~\ref{fig2_inkscape} shows a pair of connected nodes: node 1---on the left---has degree $k_1$, and node 2---on the right---has degree $k_2$. We assume that $m_1$ neighbors of node 1, other than node 2, are in the active (infected) state; similarly, $m_2$ is the number of infected neighbors, other than node 1, of node 2.

Starting from the state at the top of Fig.~\ref{fig2_inkscape}, where both node 1 and node 2 are in the susceptible state, we consider possible cycles of state-transitions for the node pair, assuming that $m_1$ and $m_2$ remain unchanged. The transition from the $\{S,S\}$ state (top of figure) to the $\{S,I\}$ state (right of figure), for example, occurs at rate $F_{k_2,m_2}$, since node 2 becomes infected while it has $m_2$ infected neighbors. The transition from $\{S,I\}$ (right) to $\{I,I\}$ (bottom of Figure) occurs at rate $F_{k_1,m_1+1}$, since node 1 is required to become infected at a time when it has $m_1+1$ infected neighbors. The other rates are derived similarly.

For microscopic reversibility, it is necessary that the product of the rates around a closed cycle are independent of the direction of rotation around the cycle \cite{Barratbook,deOliveira93,Krapivskybook}. From Fig.~\ref{fig2_inkscape}, this means we must have
\begin{equation}
F_{k_2,m_2} F_{k_1,m_1+1} R_{k_2,m_2+1} R_{k_1,m_1} = F_{k_1,m_1} F_{k_2,m_2+1} R_{k_1,m_1+1} R_{k_2,m_2}.
\end{equation}
Rearranging gives the condition
\begin{equation}
\frac{F_{k_1,m_1+1}}{F_{k_1,m_1}} \frac{R_{k_1,m_1}}{R_{k_1,m_1+1}}
=\frac{F_{k_2,m_2+1}}{F_{k_2,m_2}} \frac{R_{k_2,m_2}}{R_{k_2,m_2+1}}.
\end{equation}
Since $k_1$ and $k_2$ (and $m_1$ and $m_2$) are independent, this requires that
\begin{equation}
\frac{F_{k,m+1}}{F_{k,m}} \frac{R_{k,m}}{R_{k,m+1}} = a \quad \text{ for all }k\text{, and all }m\le k,
\end{equation}
where $a$ is a constant. Defining $u_m = F_{k,m}/R_{k,m}$, this equation can be written as the first-order difference equation
\begin{equation}
u_{m+1} = a u_m,
\end{equation}
with the solution $u_m = a^m u_0$. In terms of the rates, this means that detailed balance requires that
\begin{equation}
\frac{F_{k,m}}{R_{k,m}} = a^m \frac{F_{k,0}}{R_{k,0}}.
\end{equation}
Identifying $b_k$ as $F_{k,0}/R_{k,0}$ yields Eq.~(\ref{T}).

\section{Derivation of equation (\ref{rhoksoln})} \label{appC}
Here we manipulate Eqs.~(\ref{beqn}) and (\ref{aeqn}) for equilibrium models to produce the solution (\ref{rhoksoln}) and the algebraic equation (\ref{wsoln}) for $\overline p$. Starting with Eq.~(\ref{aeqn}), we can solve for $\overline q$, yielding
\begin{equation}
\overline q = \frac{\overline p a}{1-\overline p + \overline p a}. \label{q1}
\end{equation}
Solving (\ref{beqn}) for $\overline \rho_k$ yields
\begin{equation}
{\overline \rho}_k= \frac{b_k}{b_k+\left(\frac{1-\overline q}{1-\overline p}\right)^k} \label{rhok1}.
\end{equation}
Inserting expression (\ref{q1}) into (\ref{rhok1}) gives the desired equation (\ref{rhoksoln}).

Next, we consider the identity (\ref{star}), for which the PA ansatz gives
\begin{equation}
\sum_k P_k (1-\overline{\rho}_k) k {\overline p}_k = \sum_k P_k \overline{\rho}_k k (1-\overline{q}_k).\label{eq:55}
\end{equation}
Since $\overline{p}_k$ and $\overline{q}_k$ are independent of $k$ for the case considered here---see Eq.~(\ref{pindepk})---we obtain the simpler expression
\begin{equation}
\overline p (1-\overline \omega) = \overline \omega (1-\overline q) \label{pqw}
\end{equation}
where $\overline \omega =\left< \frac{k}{z} \overline{\rho}_k \right>$. Solving Eq.~(\ref{pqw}) for $\overline \omega$ and using Eq.~(\ref{q1}) gives
\begin{equation}
\overline \omega = \frac{\overline p(1-\overline p + \overline p a)}{1-{\overline p}^2+{\overline p}^2 a}\label{wp}
\end{equation}
and equating this to $\sum_k \frac{k}{z} P_k \overline{\rho}_k$, with Eq.~(\ref{rhoksoln}), gives Eq.~(\ref{wsoln}).

\section{Critical point of equilibrium  models with up-down symmetry} \label{appD}
Equilibrium spin models with up-down symmetry are discussed in Sec.~\ref{sec:spin}, and it is shown there that they constitute a special case of the class of models defined by relation (\ref{T}). The up-down symmetry imposes condition (\ref{barel}) on the parameters in the solution (\ref{rhoksoln}), so that $\overline{\rho}_k$ can be expressed in the form
\begin{equation}
\overline{\rho}_k = \frac{1}{1+\left( \frac{\sqrt{a}}{1-\overline p + \overline p a}\right)^{k}}. \label{rhok2}
\end{equation}
Recall from Sec.~\ref{sec:spin} that a symmetric solution of the AME with $\rho_k=1/2$ exists for all $t$; comparison with (\ref{rhok2}) shows this corresponds to the solution with $\overline p = 1/(1+\sqrt{a})$. Inserting this into the expression (\ref{wp}) for $\overline \omega$---the probability that one end of a randomly chosen edge is infected---gives $\overline \omega=1/2$ for the symmetric solution.

Next, we investigate the possibility of other solutions lying near the symmetric solution; in the language of dynamical systems, we construct the normal form of the bifurcation \cite{Strogatzbook}. We choose $\overline \omega$ as the order parameter, and let $\overline \omega = \frac{1}{2}+\epsilon$, for small $\epsilon$, to explore the neighborhood of the symmetric solution. Equation (\ref{wp}) can then be inverted to give $\overline p$ in terms of $\epsilon$, and inserting this into Eq.~(\ref{wsoln}) leads to a self-consistent equation for $\epsilon$:
\begin{equation}
\epsilon + \frac{1}{2} = \sum_k \frac{k}{z} P_k \left[1+ \left(\frac{2\epsilon+\sqrt{a+4(1-a)\epsilon^2}}{(1+2\epsilon)\sqrt{a}}\right)^k\right]^{-1}.
\end{equation}
After some rearrangement, this can be written as
\begin{equation}
\epsilon = \frac{1}{2}\sum_k \frac{k}{z} P_k \tanh\left[ -\frac{k}{2} \ln \left(\frac{2\epsilon+\sqrt{a+4(1-a)\epsilon^2}}{(1+2\epsilon)\sqrt{a}}\right) \right] \label{eps}
\end{equation}
and, in the case $a=e^{4J/T}$, this is very similar to Eq.~(13) of \cite{Dorogovtsev02} and Eq.~(28) of \cite{Leone02}, which were derived---using very different methods (recursion method and replica trick)---for the Ising model.

It follows that the structure of solutions of Eq.~(\ref{eps}) near the critical point (bifurcation point) can be analysed---for the class of  models obeying both (\ref{T}) and (\ref{T2})---using the same methods for (\ref{eps}) as used in \cite{Leone02,Dorogovtsev02}. If $P_k$ has finite fourth moment, for example, the right hand side of (\ref{eps}) can be expanded as a Taylor series in small $\epsilon$, giving
\begin{equation}
\epsilon \sim c_1 \epsilon + c_3 \epsilon^3 + \mathcal{O}(\epsilon^5)\quad \text{ as }\epsilon\to 0, \label{asyeps1}
\end{equation}
where $c_1=\frac{\left< k^2\right>}{2 z}\left(1-\frac{1}{\sqrt{a}}\right)$ and $c_3$ is a coefficient involving moments of $P_k$ up to the fourth. Equation (\ref{asyeps1}) possesses the symmetric solution $\epsilon=0$ for all parameter values, but solutions with non-zero $\epsilon$ can also be found, if leading order terms balance such that
\begin{equation}
c_1-1+c_3 \epsilon^2=0 \label{63new}.
\end{equation}
The $\epsilon$-independent coefficient $c_1-1$ in this equation is negative for small values of the parameter $a$, but it is positive when $a$ exceeds the value $a_c$ given by Eq.~(\ref{acrit}), while $c_3$ is negative at $a=a_c$. Thus Eq.~(\ref{63new}) has real-valued solutions for $\epsilon$ provided that $a\ge a_c$, and so $a_c$ marks the pitchfork bifurcation point (critical point).

The case $P_k\sim k^{-\gamma}$ can be examined in the same fashion as in \cite{Leone02,Dorogovtsev02}. The small-$\epsilon$ expansion of (\ref{eps}) has leading-order terms of the form
\begin{equation}
\epsilon \sim c_1 \epsilon + c_3 \epsilon^3+ c_0 \epsilon^{\gamma-2} \quad \text{ as }\epsilon\to 0.
\end{equation}
If $\gamma$ is in the range $2<\gamma<3$, the $\epsilon^{\gamma-2}$ term dominates both the linear and cubic terms, and the critical point is determined by the vanishing of coefficient $c_0$: this occurs at $a=a_c=1$. For exponents in the range $3<\gamma<5$, the linear term dominates the $\epsilon^{\gamma-2}$ term, and the critical point is again (\ref{acrit}), but with a different scaling near criticality. For $\gamma>5$ we recover the case (\ref{asyeps1}).

\section{PA critical point of symmetric models on $z$-regular random graphs} \label{appE}
For models with the symmetry (\ref{T2}), but not necessarily possessing the equilibrium property (\ref{T}), we focus here on the steady-state of the PA equations (\ref{PA}) on $z$-regular graphs. In particular, we suppose that the dynamics (through the rates $F_{k,m}$) depends on a parameter $Q$, and we derive the  condition (\ref{z2}) determining the critical value (bifurcation point) of this parameter.

We begin by noting that the property (\ref{T2}) means that $\overline \rho = 1/2$ is always a steady-state solution of (\ref{PA}). Using this value, and property (\ref{T2}), on the right-hand side of the first of the PA equations (\ref{PA}) gives the steady-state relation
\begin{align}
0 & = -\frac{1}{2}\sum_{m=0}^z F_{z,z-m} B_{z,m}(\overline q) + \frac{1}{2}\sum_{m=0}^z F_{z,m}B_{z,m}(\overline p) \nonumber\\
& = -\frac{1}{2} \sum_{m=0}^z F_{z,m} \left[ B_{z,m}(1-\overline q) - B_{z,m}(\overline p)\right],
\end{align}
which is satisfied---as are the other steady PA equations---if $\overline q = 1- \overline p$ in the symmetric state. Using this to replace $\overline q$ in the second of the PA equations gives, after some manipulation, the implicit relation for the steady-state value of $\overline p$:
\begin{equation}
\sum_{m=0}^z \left(1-\frac{2 m}{z}\right) F_{z,m} B_{z,m}(\overline p) = 0. \label{L31A}
\end{equation}

Next we consider the possibility of steady-states that are distinct from the symmetric state solution discussed above; we introduce the symbol $\sigma$ as a convenient shorthand for the symmetric state: $\sigma= \left\{ \overline \rho = 1/2, \overline p \text{ solves (\ref{L31A}), }\overline q=1-\overline p\right\}$.
For a general---possibly non-symmetric---steady state, the first of the PA equations is
\begin{equation}
0 = -\overline \rho \sum_{m=0}^z F_{z,m} B_{z,m}(1-\overline q) + (1-\overline \rho) \sum_{m=0}^z F_{z,m} B_{z,m}(\overline p). \label{rhoPA1}
\end{equation}
We now treat $\overline \rho$, $\overline p$, and $\overline q$ as implicit functions of $Q$, the parameter defining the rates $F_{z,m}$. Differentiating both sides of equation (\ref{rhoPA1}) with respect to $Q$, evaluating at the symmetric state $\sigma$, using the relations $\left. \overline q\right|_\sigma = 1-\left.\overline p \right|_\sigma$ and, from (\ref{pqw}),
\begin{align}
\left. \frac{\partial}{\partial Q}\overline q\right|_\sigma &= -\left. \frac{\partial}{\partial Q}\left(\frac{1-\overline \rho}{\overline \rho}\overline p\right)\right|_\sigma\nonumber\\
&= 4 \left.\frac{\partial \overline \rho}{\partial Q}\right|_\sigma \left.\overline p\right|_\sigma - \left. \frac{\partial \overline p}{\partial Q}\right|_\sigma,
\end{align}
leads eventually to the relation
\begin{equation}
0 = 2 \frac{\partial \overline \rho}{\partial Q} \left[ 1+ \frac{z}{2} \frac{2 \overline p - 1}{1-\overline p}\right] \sum_{m=0}^z F_{z,m}B_{z,m}(\overline p),
\end{equation}
where all terms are evaluated at the symmetric state $\sigma$. For this equation to be true, we must have either $\partial \overline \rho/\partial Q=0$ or the term in square brackets must vanish (note the third term is a sum of positive terms so cannot be zero). The symmetry-breaking bifurcation points correspond to the vanishing of the square bracketed term, and this gives the critical value of $\overline p$:
\begin{equation}
\overline p_c = \frac{z-2}{2 z-2}.
\end{equation}
Finally, inserting this value into the general steady-state solution (\ref{L31A}) gives the criticality condition (\ref{z2}) on the rate $F_{z,m}$ ($=R_{z,z-m}$) of the PA dynamics.

\section{Derivation of reduced-dimension equations for threshold models}\label{appF}
Our goal here is to
 demonstrate that Eqs.~(\ref{2}) and (\ref{1}) reduce to Eqs.~(\ref{3})--(\ref{5}) through an exact solution of (\ref{1}) given by (\ref{6}).

We proceed to insert the ansatz (\ref{6}) into the AME (\ref{1}) for $m<M_\kv$ (for which $m$ values we have $F_{k,m}=0$). The left hand side of (\ref{1}) then gives, after some rearrangement,
\begin{equation}
\dot{s}_{\kv,m} = \left(1- \rho_\kv(0)\right) \left[ \frac{m}{\phi} - \frac{k-m}{1-\phi}\right] B_{k,m}(\phi)\, \dot \phi,
\end{equation}
where dots denote time derivatives.
Using the identity
\begin{equation}
B_{k,m-1}(\phi) = \frac{1-\phi}{\phi} \frac{m}{k-m+1} B_{k,m}(\phi)
\end{equation}
on the right hand side of Eq.~(\ref{1})  yields the condition
\begin{equation}
\dot \phi = \beta^s (1-\phi) \label{7}
\end{equation}
on the function $\phi(t)$ for the ansatz (\ref{6}) to be an exact solution of (\ref{1}). From the initial condition on $s_{\kv,m}$ we also obtain
\begin{equation}
\phi(0)=\rho(0) = \sum_\kv \frac{k}{z} P_\kv \rho_\kv(0). \label{8}
\end{equation}
Comparing (\ref{3}) and (\ref{7}) we see that it remains for us to show that
\begin{equation}
\beta^s = \frac{g(\phi)-\phi}{1-\phi}. \label{9}
\end{equation}
We first prove a preliminary and rather general result, as follows. Multiplying the AME (\ref{1}) by $(k-m)P_\kv$, summing over $m=0,\ldots,k$ and over the $\kv$-classes gives:
\begin{eqnarray}
\frac{d}{dt} \sum_\kv P_\kv \sum_m (k-m)s_{\kv,m} &=& - \sum_\kv P_\kv \sum_m F_{\kv,m}(k-m) s_{\kv,m} \nonumber\\
&& \hspace{-0.5cm} - \beta^s \sum_\kv P_\kv \sum_m \left[ (k-m)^2 s_{\kv,m} - (k-m)(k-m+1) s_{\kv,m-1}\right] \label{10}.
\end{eqnarray}
Note that the second term on the right hand side is a telescoping series that reduces to
 \begin{equation}
 -\beta^s \sum_\kv P_\kv \sum_m (k-m) s_{\kv,m}.
 \end{equation}
 and that the definition of $\beta^s$ enables us to express the first term on the right hand side as
 \begin{equation}
 -\beta^s \sum_\kv P_\kv \sum_m (k-m) s_{\kv,m}. \label{11}
 \end{equation}
 Therefore (\ref{10}) can be solved for $\beta^s$ to yield
 \begin{eqnarray}
 \beta^s &=& \frac{-\frac{1}{2} \frac{d}{dt} \sum_\kv P_\kv \sum_m (k-m) s_{\kv,m}}{\sum_\kv P_\kv \sum_m (k-m) s_{\kv,m}} \nonumber \\
 &=& -\frac{1}{2} \frac{d}{d t} \ln \left( \sum_\kv P_\kv \sum_m (k-m) s_{\kv,m}\right) \label{12}.
 \end{eqnarray}

 Rewriting (\ref{7}) as $\beta^s = -\frac{d}{dt}\ln(1-\phi)$ and comparing with (\ref{12}) gives the result that $\sum_\kv P_\kv \sum_m (k-m)s_{\kv,m}$ and $(1-\phi)^2$ are equal, up to a multiplicative constant.
 Using the initial conditions on $s_{\kv,m}$ and $\phi$ determines the constant, giving
 \begin{equation}
 \sum_\kv P_\kv \sum_m (k-m) s_{\kv,m} = z (1-\phi)^2 \label{13}.
 \end{equation}

We now use this result and the definition of $\beta^s$ to prove (\ref{9}). For the infection probabilities (\ref{2}), $\beta^s$ is given by
\begin{eqnarray}
\beta^s &=& \frac{\sum_\kv P_\kv \sum_{m\ge M_\kv} (k-m) s_{\kv,m}}{\sum_\kv P_\kv \sum_m (k-m) s_{\kv,m}} \nonumber\\
&=& \frac{{\sum_\kv P_\kv \sum_m (k-m) s_{\kv,m}}-{\sum_\kv P_\kv \sum_{m<M_\kv}(k-m) s_{\kv,m}} }{\sum_\kv P_\kv \sum_m (k-m) s_{\kv,m}} \nonumber\\
 &=& \frac{z(1-\phi)^2 - {\sum_\kv P_\kv \sum_{m<M_\kv}(k-m)(1-\rho_\kv(0)) B_{k,m}}}{z(1-\phi)^2} \label{14},
 \end{eqnarray}
 where the final line uses (\ref{13}) twice and the ansatz (\ref{6}). Dividing the numerator and denominator by $z(1-\phi)$ and using the identities $(k-m)B_{k,m}(\phi)=k(1-\phi)B_{k-1,m}(\phi)$ and $\sum_{m<M_\kv} B_{k-1,m}(\phi) = 1 - \sum_{m\ge M_\kv} B_{k-1,m}(\phi)$ leads to
 \begin{equation}
\beta^s = \frac{g(\phi)-\phi}{1-\phi},
\end{equation}
as claimed.

Finally, let's show that the fraction $\rho(t)$ of active nodes given in the AME as
$ \rho(t)=1-\sum_\kv P_\kv \sum_m s_{\kv,m}$, obeys equations (\ref{3})--(\ref{4}). Multiplying equation (\ref{1}) by $P_\kv$ and summing over $m$ and over $\kv$-classes gives
\begin{equation}
\sum_\kv P_\kv \sum_m \dot{s}_{\kv,m} = - \sum_\kv P_\kv \sum_m F_{\kv,m} s_{\kv,m} - \beta^s \sum_\kv P_\kv \sum_m \left[(k-m) s_{\kv,m}- (k-m+1)s_{\kv,m-1}\right],
\end{equation}
where the second term on the right hand side is easily seen to telescope to zero. Thus we have
\begin{eqnarray}
\dot \rho(t)  &=& \sum_\kv P_\kv \sum_m F_{\kv,m} s_{\kv,m} \nonumber\\
&=& \sum_\kv P_\kv \sum_m s_{\kv,m} - \sum_\kv P_\kv \sum_{m<M_\kv} s_{\kv,m} \nonumber\\
&=& 1-\rho - \sum_\kv P_\kv (1-\rho_\kv(0))\left[1-\sum_{m\ge M_\kv} B_{k,m}(\phi)\right],
\end{eqnarray}
using (\ref{6}). It is straightforward to verify that this reduces to equations (\ref{3})--(\ref{4}).

For completeness, note that the distribution of $s_{\kv,m}$ for $m \ge M_\kv$ is, in general, not of the binomial form (\ref{6}), which applies only to $m$ values below the threshold $M_\mathbf{k}$. To obtain the values of $s_{\kv,m}$ for these $m$ values, note that equation (\ref{1}) has a solution giving $s_{\kv,m}$ explicitly
in terms of $s_{\kv,m-1}$ by using the integrating factor
\begin{equation}
\exp\left( t + (k-m) \int \beta^s dt\right) = e^t (1-\phi)^{-(k-m)},
\end{equation}
and using (\ref{7}) for $\beta^s$. Then we have
\begin{eqnarray}
s_{\kv,m}(t) &=& s_{\kv,m}(0)\,e^{-t}\left(\frac{1-\phi(t)}{1-\phi(0)}\right)^{k-m} \nonumber\\
&& \hspace{-0.2cm}+(k-m+1)\,e^{-t}(1-\phi)^{k-m} \int_0^t e^\tau\left(1-\phi(\tau)\right)^{-(k-m)} \beta^s(\tau) s_{\kv,m-1}(\tau)\, d \tau
\end{eqnarray}
which can be solved recursively for $m=M_\kv+1$ to $m=k$.


\begin{thebibliography}{99}
\bibitem{Castellano09} C.~Castellano, S.~Fortunato, and V.~Loreto, \emph{Statistical Physics of Social Dynamics}, {Rev. Mod. Phys.} \textbf{81}, 591  (2009).
\bibitem{Castellano12a} C.~Castellano, \emph{Social Influence and the Dynamics of
Opinions: The Approach of
Statistical Physics}, {Manage. Decis. Econ.} \textbf{33}, 311 (2012).
\bibitem{Barratbook} A.~Barrat, M.~Barth\'{e}lemy, A.~Vespignani, \emph{Dynamical Processes on Complex Networks}, (Cambridge University Press, Cambridge, 2008).
\bibitem{Newmanbook} M.~E.~J.~Newman, \emph{Networks: An Introduction} (Oxford University Press, Oxford, 2010).
\bibitem{Liggettbook} T.~M.~Liggett, \emph{Interacting Particle Systems} (Springer, New York, 1985).
\bibitem{Sood05} V.~Sood and S.~Redner, \emph{Voter Model on Heterogeneous Graphs}, {Phys. Rev. Lett.} \textbf{94}, 178701 (2005).
\bibitem{deOliveira92} M.~J.~de~Oliveira, \emph{Isotropic Majority-Vote Model on a
Square Lattice}, {J. Stat. Phys.} \textbf{66}, 273 (1992).
\bibitem{Pereira05} L.~F.~C.~Pereira and F.~G.~Brady  Moreira, \emph{Majority-vote Model on Random Graphs}, {Phys. Rev. E.} \textbf{71}, 016123 (2005).
\bibitem{PastorSatorras01} R.~Pastor-Satorras and A.~Vespignani, \emph{Epidemic Spreading in Scale-Free Networks}, {Phys. Rev. Lett.} \textbf{86}, 3200 (2001).
\bibitem{Hill10} A.~L.~Hill, D.~G.~Rand, M.~A.~Nowak and N.~A.~Christakis, \emph{Infectious Disease Modeling of Social Contagion in
Networks}, {PLoS Comput. Biol.} \textbf{6}, e1000968 (2010).
\bibitem{Young09} H.~P.~Young, \emph{Innovation Diffusion in Heterogeneous Populations:
Contagion, Social Influence, and Social Learning}, {Am. Econ. Rev.} \textbf{99}, 1899 (2009).
\bibitem{Granovetter78} M.~Granovetter, \emph{Threshold Models of Collective Behavior}, {Am. J. Sociol.} \textbf{83}, 1420 (1978).
\bibitem{Watts02} D.~J.~Watts, \emph{A Simple Model of Global Cascades on
Random Networks}, {Proc. Nat. Acad. Sci. USA} \textbf{99}, 5766 (2002).
\bibitem{Centola07} D.~Centola, V.~M.~Egu\'{i}luz, and M.~W.~Macy, \emph{Cascade Dynamics of Complex Propagation}, {Physica A} \textbf{374}, 449 (2007).
\bibitem{Dodds12} P.~S.~Dodds, K.~D.~Harris, and C.~M.~Danforth, \emph{Limited Imitation Contagion on Random Networks: Chaos, Universality, and
Unpredictability}, arXiv:1208.0255 (2012).
\bibitem{Romero11} D.~M.~Romero, B.~Meeder, and J.~Kleinberg, \emph{Differences in the Mechanics of Information Diffusion
Across Topics: Idioms, Political Hashtags, and Complex
Contagion on Twitter}, {Proc.~20th international conference on World Wide Web}, 695 (ACM, 2011).
\bibitem{VerSteeg11} G.~Ver~Steeg, R.~Ghosh, and K.~Lerman, \emph{What Stops Social Epidemics?}, {Proc. Fifth International AAAI Conference on Weblogs and Social Media}, arXiv:1102.1985  (2011).
\bibitem{Kleinbergbook} D.~Easley and J.~Kleinberg, \emph{Networks, Crowds, and Markets} (Campbridge University Press, New York, 2010).
\bibitem{GonzalezBailon11} S.~Gonzalez-Bailon, J.~Borge-Holthoefer, A.~Rivero, and Y.~Moreno, \emph{The Dynamics of Protest Recruitment
through an Online Network}, {Scientific Reports} \textbf{1}, 197 (2011).
\bibitem{Bakshy11} E.~Bakshy, J.~M.~Hofman, W.~A.~Mason and D.~J.~Watts, \emph{Everyone's an Influencer:
Quantifying Influence on Twitter}, {Proc.~4th ACM international conference on Web search and data mining}, 65 (ACM, 2011).
\bibitem{Hodas12} N.~O.~Hodas and K.~Lerman, \emph{How Visibility and Divided Attention Constrain
Social Contagion}, arXiv:1205.2736 (2012).
\bibitem{Centola10} D.~Centola, \emph{The Spread of Behavior in an Online Social Network Experiment}, {Science} \textbf{329}, 1194 (2010).
\bibitem{Shao09} J.~Shao, S.~Havlin, and H.~E.~Stanley, \emph{Dynamic Opinion Model and Invasion Percolation}, {Phys. Rev. Lett.} \textbf{103}, 018701 (2009).
\bibitem{PerezReche11} F.~J.~P\'{e}rez-Reche, J.~J.~Ludlam, S.~N.~Taraskin, and C.~A.~Gilligan, \emph{Synergy in Spreading Processes: From Exploitative to Explorative Foraging Strategies}, {Phys. Rev. Lett.} \textbf{106}, 218701 (2011).
\bibitem{Dorogovtsev02} S.~N.~Dorogovtsev, A.~V.~Goltsev, and J.~F.~F.~Mendes, \emph{Ising Model on Networks with an Arbitrary Distribution of Connections}, {Phys. Rev. E.} \textbf{66}, 016104 (2002).
\bibitem{Leone02} M.~Leone, A.~V\'{a}zquez, A.~Vespignani and R.~Zecchina, \emph{Ferromagnetic Ordering in Graphs with Arbitrary Degree
Distribution}, {Eur. Phys. J. B} \textbf{28}, 191 (2002).
\bibitem{Galam82} S.~Galam, Y.~Gefen, and Y.~Shapir, \emph{Sociophysics: A Mean Behavior Model for the Process of Strike}, {J. Math. Sociol.} \textbf{9}, 1 (1982).
\bibitem{Castellano12b} C.~Castellano and R.~Pastor-Satorras, \emph{Universal and Nonuniversal Features of the Generalized Voter Class for Ordering
Dynamics in Two Dimensions}, {Phys. Rev. E} \textbf{86}, 051123 (2012).
\bibitem{Vazquez08b} F.~Vazquez and C.~L\'opez, \emph{Systems with Two Symmetric Absorbing States: Relating the Microscopic Dynamics
with the Macroscopic Behavior}, {Phys. Rev. E}, \textbf{78}, 061127 (2008).
\bibitem{footnote1}{Dynamical correlations should not be confused with degree-degree (structural) correlations, which are a property of the network rather than of the dynamics. In this paper we assume zero structural correlations in the networks (uncorrelated configuration model \cite{Bender78,Bollobas80,Newmanbook}).}
\bibitem{Bender78} E.~A.~Bender and E.~R.~Canfield, \emph{The Asymptotic Number of Labeled Graphs with Given Degree Sequences}, {J. Comp. Theory Ser. A} \textbf{24}, 296 (1978).
\bibitem{Bollobas80} B.~Bollob\'{a}s, \emph{A Probabilistic Proof of an Asymptotic Formula for the Number of Labelled Random Graphs}, {Eur. J. Comb.} \textbf{1}, 311 (1980).
\bibitem{Castellano06} C.~Castellano and R.~Pastor-Satorras, \emph{Zero Temperature Glauber Dynamics on
Complex Networks}, {J. Stat. Mech.} P05001 (2006).
\bibitem{Gleeson12} J.~P.~Gleeson, S.~Melnik, J.~A.~Ward, M.~A.~Porter, and P.~J.~Mucha, \emph{Accuracy of Mean-Field Theory for Dynamics on Real-World Networks}, {Phys. Rev. E.} \textbf{85}, 026106 (2012).
\bibitem{Melnik11} S.~Melnik, A.~Hackett, M.~A.~Porter, P.~J.~Mucha, and J.~P.~Gleeson, \emph{The Unreasonable Effectiveness of Tree-Based Theory for Networks with Clustering}, {Phys. Rev. E.} \textbf{83}, 036112 (2011).
\bibitem{Eames02} K.~T.~D.~Eames and M.~J.~Keeling, \emph{Modeling Dynamic and Network Heterogeneities in
the Spread of Sexually Transmitted Diseases}, {Proc. Nat. Acad. Sci. USA} \textbf{99}, 13330 (2002).
\bibitem{Levin96} S.~A.~Levin and R.~Durrett, \emph{From Individuals to Epidemics}, {Phil. Trans. R. Soc. Lond. B} \textbf{351}, 1615 (1996).
\bibitem{deOliveira93} M.~J.~de~Oliveira, J.~F.~F.~Mendes, and M.~A.~Santos, \emph{Non-Equilibrium Spin Models with Ising Universal Behaviour}, {J. Phys. A: Math. Gen.} \textbf{26}, 2317 (1993).
\bibitem{Taylor12b} M.~Taylor, P.~L.~Simon, D.~M.~Green, T.~House, and I.~Z.~Kiss, \emph{From Markovian to Pairwise Epidemic Models
and the Performance of Moment Closure Approximations}, {J. Math. Biol.} \textbf{64}, 1021 (2012).
\bibitem{Vazquez08} F.~Vazquez and V.~M.~Eguíluz, \emph{Analytical Solution of the Voter Model on
Uncorrelated Networks}, {New. J. Phys.} \textbf{10}, 063011 (2008).
\bibitem{Vazquez10} F.~Vazquez, X.~Castell\'{o}, and M.~San~Miguel, \emph{Agent Based Models of Language
Competition: Macroscopic Descriptions
and Order–Disorder Transitions}, {J. Stat. Mech.} P04007 (2010).
\bibitem{Schweitzer09} F.~Schweitzer and L.~Behera, \emph{Nonlinear Voter Models: the Transition From Invasion
to Coexistence}, {Eur. Phys. J. B} \textbf{67}, 301 (2009).
\bibitem{Dickman86} R.~Dickman, \emph{Kinetic Phase Transitions in a Surface-Reaction Model: Mean-Field Theory}, {Phys. Rev. A.} \textbf{34}, 4246 (1986).
\bibitem{Marceau10} V.~Marceau, P.-A.~No\"{e}l, L.~H\'{e}bert-Dufresne, A.~Allard, and L.~J.~Dub\'{e}, \emph{Adaptive Networks: Coevolution of Disease and Topology}, {Phys. Rev. E} \textbf{82}, 036116 (2010).
\bibitem{Lindquist11} J.~Lindquist, J.~Ma, P.~van~den~Driessche, and F.~H.~Willeboordse, \emph{Effective Degree Network Disease Models}, {J. Math. Biol.} \textbf{62}, 143 (2011).
\bibitem{Gleeson11} J.~P.~Gleeson, \emph{High-Accuracy Approximation of Binary-State Dynamics on Networks}, {Phys. Rev. Lett.} \textbf{107}, 068701 (2011).
\bibitem{Petermann04} T.~Petermann and P.~De~Los~Rios, \emph{Cluster Approximations for Epidemic Processes:
a Systematic Description of Correlations Beyond the Pair Level}, {J. Theor. Biol.}, \textbf{229}, 1 (2004).
\bibitem{Durrett12} R.~Durrett, J.~P.~Gleeson, A.~L.~Lloyd, P.~M.~Mucha, F.~Shi, D.~Sivakoff, J.~E.~S.~Socolar and C.~Varghese, \emph{Graph fission in an evolving voter model}, {Proc. Nat. Acad. Sci. USA} \textbf{109}, 3682 (2012).
\bibitem{Taylor12a} M.~Taylor, T.~J.~Taylor, and I.~Z.~Kiss, \emph{Epidemic Threshold and Control in a Dynamic Network}, {Phys. Rev. E.} \textbf{85}, 016103 (2012).
\bibitem{Newman09} M.~E.~J.~Newman, \emph{Random Graphs with Clustering}, {Phys. Rev. Lett.} \textbf{103}, 058701 (2009); J.~C.~Miller, \emph{Percolation and Epidemics in Random Clustered Networks}, {Phys. Rev. E} \textbf{80}, 020901(R) (2009).
\bibitem{Gleeson09}    J.~P.~Gleeson, \emph{Bond Percolation on a Class of Clustered Random Networks}, {Phys. Rev. E} \textbf{80}, 036107 (2009).
\bibitem{Hackett11} A.~Hackett, S.~Melnik, and J.~P.~Gleeson, \emph{Cascades on a Class of Clustered Random Networks}, {Phys. Rev. E}, \textbf{83}, 056107 (2011).
\bibitem{HebertDufresne10} L.~H\'{e}bert-Dufresne, P.-A.~No\"{e}l, V.~Marceau,   A.~Allard, and L.~J.~Dub\'{e}, \emph{Propagation Dynamics on Networks Featuring Complex Topologies}, {Phys. Rev. E}, \textbf{82}, 036115 (2010).
\bibitem{footnote2}{The case of symmetric  models, where the dynamics are unchanged by simultaneous flipping of all nodes' spins, is considered in detail in Sec.~\ref{sec:spin}.}
\bibitem{Baileybook} N.~T.~J.~Bailey, \emph{The Mathematical Theory of Infectious Diseases} (Griffin, London, 1975).
\bibitem{Andersonbook} R.~M.~Anderson and R.~M.~May, \emph{Infectious Diseases in Humans} (Oxford University Press, Oxford, 1992).
\bibitem{Harris74} T.~E.~Harris, \emph{Contact Processes on a Lattice}, {Ann. Probability} \textbf{2}, 969 (1974).
\bibitem{Bass69} F.~Bass, \emph{A New Product Growth Model for Conmsumer Durables}, {Management Science} \textbf{15}, 215 (1969).
\bibitem{Dodds04}  P.~S.~Dodds and D.~J.~Watts, \emph{Universal Behavior in a Generalized Model of Contagion}, {Phys. Rev. Lett.} \textbf{92}, 218701 (2004).
\bibitem{Watts07}  D.~J.~Watts and P.~S.~Dodds, \emph{Influentials, Networks, and Public Opinion
Formation}, {Journal of Consumer Research} \textbf{34}, 441 (2007).
\bibitem{Kirman93} A.~Kirman, \emph{Ants, Rationality, and Recruitment}, {The Quarterly Journal of Economics} \textbf{108}, 137 (1993).
\bibitem{Alfarano05} S.~Alfarano, T.~Lux, and F.~Wagner, \emph{Estimation of Agent-Based Models:
The Case of an Asymmetric Herding Model}, {Comp. Economics} \textbf{26}, 19 (2005).
\bibitem{Gontis12} V.~Gontis, A.~Kononovicius, and S.~Reimann, \emph{The class of Nonlinear Stochastic Models as a Background for the Bursty Behavior in Financial Markets}, {Advs. Complex Syst.} \textbf{15}, 1250071 (2012).
\bibitem{Molofsky99} J.~Molofsky, R.~Durrett, J.~Dushoff, D.~Griffeath and S.~Levin, \emph{Local Frequency Dependence and
Global Coexistence}, {Theor. Pop. Biol.} \textbf{55}, 270 (1999).
\bibitem{Abrams03} D.~M.~Abrams and S.~H.~Strogatz, \emph{Modelling the Dynamics of Language Death}, {Nature} \textbf{424} 900 (2003).
\bibitem{Krapivskybook} P.~L.~Krapivsky, S.~Redner, and E.~Ben-Naim, \emph{A Kinetic View of Statistical Physics} (Campbridge University Press, New York, 2010).
\bibitem{Glauber63} R.~J.~Glauber, \emph{Time-Dependent Statistics of the Ising Model}, {J. Math. Phys.} \textbf{4}, 294 (1963).
\bibitem{Metropolis53} N.~Metropolis, A.~W.~Rosenbluth, M.~N.~Rosenbluth and A.~H.~Teller, \emph{Equation of State Calculations by Fast Computing Machines}, {J. Chem. Phys.} \textbf{21}, 1087 (1953).
\bibitem{Gleeson07} J.~P.~Gleeson and D.~J.~Cahalane, \emph{Seed Size Strongly Affects Cascades on Random Networks}, {Phys. Rev. E.} \textbf{75}, 056103 (2007).
\bibitem{Morris00} S.~Morris, \emph{Contagion}, {Review of Economic Studies} \textbf{67} 57 (2000).
\bibitem{Montanari10} A.~Montanari and A.~Saberi, \emph{The Spread of Innovations in Social Networks}, {Proc. Nat. Acad. Sci. USA} \textbf{107}, 20196 (2010).
\bibitem{AME_solve} Octave/Matlab  files for implementing and solving the AME, PA and MF systems in (Eqs.~(\ref{seqns})--(\ref{MF})) are available for download from the author's webpage \texttt{www.ul.ie/gleesonj}. The documentation includes commands to reproduce the results shown in the figures of this paper, and in \cite{Gleeson11}.
\bibitem{footnote3}{The case where $F_{k,m}=0$ with some $R_{k,m}$ nonzero is obviously equivalent, up to swapping of state labels, to the case considered here, and so is not considered separately.}    
\bibitem{footnote4}{The networks used in numerical simulations are sufficiently large to ensure that there are multiple nodes of every degree class considered in the AME: this avoids finite-size effects due to having small numbers of nodes described by a population-level quantity such as $\rho_k(t)$.}
\bibitem{Strogatzbook} S.~H.~Strogatz, \emph{Nonlinear Dynamics and Chaos} (Perseus Books, 2000).
\bibitem{Dorogovtsev08} S.~N.~Dorogovtsev, A.~V.~Goltsev, and J.~F.~F.~Mendes, \emph{Critical Phenomena in Complex Networks}, {Rev. Mod. Phys.} \textbf{80}, 1275  (2008).
\bibitem{Drouffe99} J-M.~Drouffe and C.~Godr\`{e}che, \emph{Phase Ordering and Persistence in a Class of Stochastic
Processes Interpolating Between the Ising and Voter Models}, {J. Phys. A: Math. Gen.} \textbf{32}, 249 (1999).
\bibitem{Galstyan07} A.~Galstyan and P.~Cohen, \emph{Cascading Dynamics in Modular Networks}, {Phys. Rev. E.} \textbf{75}, 036109 (2007).
\bibitem{footnote5}{The extreme case $dt=1$ is called \emph{synchronous updating} \cite{Gomez10,Dodds12,Gleeson08}; however, our interest is in the asynchronous-updating case with $dt$ infinitesimally small, which gives continuous-time dynamics in the limit $dt \to 0$.}
\bibitem{Gomez10} S.~G\'{o}mez, A.~Arenas, J.~Borge-Holthoefer, S.~Meloni, and Y.~Moreno, \emph{Discrete-Time Markov Chain Approach to Contact-Based Disease Spreading in Complex
Networks}, {Europhys. Lett.} \textbf{89}, 38009 (2010).
\bibitem{Gleeson08} J.~P.~Gleeson, \emph{Cascades on Correlated and Modular Random Networks}, {Phys. Rev. E.} \textbf{77}, 046117 (2008).
 \bibitem{footnote6}{Chapter 19 of \cite{Kleinbergbook} shows how a coordination game played on a network---as a model of direct-benefit effects from the diffusion of new behavior \cite{Morris00}---can also be expressed as a threshold model of the type considered here.}
\bibitem{Hackett_thesis11} A.~W.~Hackett, \emph{Cascade Dynamics on Complex Networks}, PhD thesis, University of Limerick (2011).
\bibitem{Noel12} P.~A.~No\"{e}l, A.~Allard,    L.~H\'{e}bert-Dufresne, V.~Marceau, and L.~J.~Dub\'{e}, \emph{Propagation on Networks: An Exact Alternative Perspective}, {Phys. Rev. E} \textbf{85}, 031118 (2012).
\bibitem{footnote7}{The main approximation here is to assume that the edge-state transition rate $\beta^s$ is the same for all $S$-$S$ edges in the network, regardless of their local neighborhood---the same assumption is made for the other rates $\gamma^s$, $\beta^i$, and $\gamma^i$. See also the explanation in \cite{Marceau10} for the SIS case.}
\bibitem{Noel09}  P.-A.~No\"{e}l, B.~Davoudi,~R.~C.~Brunham, L.~J.~Dub\'{e}, and B.~Pourbohloul, \emph{Time Evolution of Epidemic Disease on Finite and Infinite Networks}, {Phys. Rev. E} \textbf{79}, 026101 (2009).







\end{thebibliography}
\end{document}